\documentclass[11pt]{article}
\usepackage{hyperref}
\usepackage{amsmath, amsfonts, amssymb, mathrsfs, bbm}
\usepackage{dcolumn}
\usepackage{caption}
\usepackage{subcaption}
\usepackage{filemod}
\usepackage{natbib}
\usepackage[ruled, vlined]{algorithm2e}
\usepackage{floatrow}
\usepackage{setspace}
\usepackage{verbatim}
\usepackage{graphicx}
\usepackage{graphbox}

\hypersetup{
    colorlinks=true,
    linkcolor=blue, 
    urlcolor=black,
    citecolor=blue, 
    }

\title{\vspace{-0.5cm}
    Two-stage Design for Failure Probability Estimation \\
    with Gaussian Process Surrogates}
\author{Annie S. Booth\thanks{Corresponding author: Department of Statistics, 
	Virginia Tech, {\tt annie\_booth@vt.edu}} \and S. Ashwin 
	Renganathan\thanks{Department of Aerospace Engineering, Penn State}}
\date{\today}

\begin{document}

\maketitle
\bigskip

\begin{abstract} 
We tackle the problem of quantifying failure probabilities for expensive 
deterministic computer experiments with stochastic inputs under a fixed budget.  
The computational cost of the computer simulation prohibits direct Monte Carlo (MC) and 
necessitates a surrogate model, which may facilitate
either a ``surrogate MC'' estimator or a surrogate-informed importance sampling
estimator.  We embrace the former, finding importance sampling too variable
when budgets are limited, and propose a novel design strategy to effectively train
a surrogate for the purpose of failure probability estimation.  Existing works 
exhaust the entire evaluation budget on active learning through sequential 
contour location (CL), attempting to balance exploration
with exploitation of the failure contour throughout the design, but we find exhaustive CL to be
suboptimal.  Instead we propose a novel two-stage surrogate design.  In Stage 1,
we conduct sequential CL to locate the failure contour.  In Stage 2, once surrogate
learning has saturated, we use a solely exploitative strategy -- allocating the remaining 
evaluation budget to MC samples with the highest classification entropy to ensure
they are classified correctly.  We propose
a stopping criterion to determine the transition between stages without any tuning.  
Our two-stage design outperforms alternatives, including exhaustive CL and 
surrogate-informed importance
sampling, on a variety of benchmark exercises.  With these tools, 
we are able to effectively estimate small failure probabilities with only hundreds 
of simulator evaluations, showcasing functionality with both shallow and deep 
Gaussian process surrogates, and deploying our method on a simulation of 
fluid flow around an airfoil.
\end{abstract}


\section{Introduction}\label{sec:intro}

Quantifying the probability of failure in complex systems is of great
interest but is challenging due to the computational expense of the so-called
``computer experiment.'' Consider a deterministic black-box computer simulation,
$f:\mathcal{X}\rightarrow\mathbb{R}$ with inputs $\mathcal{X}\subset\mathbb{R}^d$
governed by known measure $\mathbb{P}_\mathbf{x}$ (with density $p(\mathbf{x})$).
The response is thresholded such that $f(\mathbf{x})>t$ indicates failure.  The
research objective is to estimate the failure probability,  
\begin{equation}\label{eq:problem}
\alpha = \int\mathbbm{1}_{\left\{f(\mathbf{x})>t\right\}} \, d\mathbb{P}_\mathbf{x}
= \int_{\mathbf{x}\in\mathcal{X}}\mathbbm{1}_{\left\{f(\mathbf{x})>t\right\}}\,p(\mathbf{x})\, 
    d\mathbf{x},
\end{equation}
from a small, fixed budget of simulator evaluations.

We are particularly motivated by aeronautic applications where system failures may incur
immense financial or environmental costs, or even loss of life.  In aircraft design
and development, computer simulation experiments are a natural, often mandatory, precursor 
to direct experimentation and prototyping.  For example, $f(\cdot)$ may represent the vibrations 
experienced by an aircraft which become unsafe above certain levels, or 
the fuel efficiency of an aircraft, which if too low, can be detrimental to the environment, 
obliterate fuel budgets, or overly restrict travel distances.  Inputs $\mathcal{X}$ may encapsulate
design parameters and operating conditions (such as wing shapes, 
flight speeds, and angles of attack).  Yet we acknowledge a slew of other relevant 
applications where quantifying failure probabilities
for complex simulations is critical, including structural reliability analysis
using finite element simulations \citep[e.g.,][]{mia2017modal}, Boltzmann modeling of 
chemical processes in fuel filtration \citep[e.g.,][]{belot2021impact}, and 
thermodynamic flows for nuclear reactors \citep[e.g.,][]{batet2014modelling}.

A rudimentary approach to solve Eq.~(\ref{eq:problem}) involves 
brute force Monte Carlo (MC): sample a large number of inputs, 
evaluate the computer simulation at each of these
configurations, and report the proportion that resulted in failure.  
But this approach is prohibitively expensive if the computer simulation
is costly and $\alpha$ is small, as in our target application.
In the face of expensive data, statistical models ($\hat{f}$, called ``surrogates'' or 
``emulators'') are essential.  Surrogates are trained on a limited set of simulator 
evaluations in order to provide predictions with
appropriate uncertainty quantification (UQ) at unobserved inputs.  Gaussian 
processes (GPs) are the canonical choice \citep{santner2003design,gramacy2020surrogates},
but deep Gaussian processes \citep[DGPs;][]{damianou2013deep} are becoming increasingly popular,
particularly for nonstationary computer experiments \citep{booth2024nonstationary}.  
Notably, our contribution is agnostic to surrogate choice as long as effective UQ
is provided, and is thus applicable to traditional GPs and Bayesian DGPs 
\citep{sauer2023active} and, potentially, Bayesian neural networks \citep{mackay1995bayesian}.

We briefly digress to acknowledge some early methods of failure probability 
estimation, namely the first and second order reliability methods 
\citep[FORM/SORM;][]{haldar1995first}, which leveraged 
simplistic linear approximations of the complex model
and placed all failure probability estimation in the hands of the ``most 
probable point'' (MPP).  Although advancements to FORM and 
SORM have been made with more flexible GP models 
\citep[e.g.,][]{zhang2015efficient,su2017gaussian}, 
sole reliance on a MPP is too restricting for our target applications.

The utility of a surrogate boils down to two things: how it is trained and 
how it is used. To solve Eq.~(\ref{eq:problem}), surrogate accuracy in classifying failures is of 
primary importance.  Not all training designs are created equally.  Space-filling 
designs \citep{joseph2016space} may miss failure
regions altogether, particularly when training budgets are limited.  Instead,
it is common to train a surrogate on a small initial design then pair it with a relevant
acquisition function to select subsequent inputs for evaluation through the 
simulator -- a process called ``active learning'' (AL).  The goal is to maximize
learning from limited evaluations of the expensive simulator.  
Contour location (CL) is an AL variant which targets learning of a failure contour 
$\{\mathbf{x}\in\mathcal{X} \mid f(\mathbf{x}) = t\}$.  CL acquisition functions seek to 
balance exploitation and exploration by leveraging both predicted distance from the contour and
posterior uncertainty.  Popular variations fall under expected improvement
frameworks \citep{bichon2008efficient,ranjan2008sequential}, stepwise uncertainty
reduction frameworks \citep{bect2012sequential,chevalier2014fast,azzimonti2021adaptive},
or both \citep{duhamel2023version}.  Others utilize pass/fail classification entropy 
\citep{marques2018contour,cole2023entropy}.  \citet{booth2024contour} recently
proposed a CL sequential design for DGP surrogates.
For small failure regions, contour
location may be replaced with subset simulation \citep{au2001estimation} which initially
targets a conservative threshold ($t^\star < t$) and iteratively refines it 
until the true threshold is found ($t^\star\rightarrow t$)
\citep{dubourg2011reliability,huang2016assessing,hristov2019adaptive}.  Yet these approaches
require Markov chain Monte Carlo (MCMC) sampling of cascading input distributions, 
which we intentionally avoid here.

A well-trained surrogate may be leveraged directly within an MC estimator
with surrogate predictions (which are of negligible cost) in place of true simulator 
evaluations, which we term ``surrogate MC.''  One could reasonably use the 
entire budget of simulator evaluations for contour location, then deploy the
trained surrogate in such a ``surrogate MC'' estimator.
Although this is common \citep[e.g.,][]{zhu2016reliability,cheng2020structural,li2021ass,lu2023agp},
we find that surrogate learning of failure contours quickly plateaus, and that
common surrogates fail to achieve sufficient classification accuracy 
under tightly restricted evaluation budgets.

Another common approach splits the budget of simulator evaluations into two disjoint stages:
some evaluations are used to train a surrogate and some are used for importance sampling 
\citep[IS;][]{srinivasan2002importance,tokdar2010importance}.  
IS hinges on the construction of a bias distribution 
that intentionally focuses density in the failure region.  Inputs are sampled 
from the bias distribution and are then utilized in a weighted average calculation
that corrects for the disparity between the bias density and the original 
input density.  
Identifying an effective bias distribution in the face of limited
training data is a hefty task \citep{tabandeh2022review}.  Adaptive methods abound 
\citep[e.g.,][]{oh1992adaptive,kurtz2013cross,dalbey2014gaussian,bugallo2017adaptive,cheng2023rare}.
\citet{peherstorfer2016multifidelity} propose what they termed ``multifidelity importance sampling,''
although we prefer the term ``surrogate-informed importance
sampling'' (SIIS) to avoid confusion with multifidelity surrogate modeling \citep[e.g.,][]{park2017remarks}.
In SIIS, a trained surrogate is used to identify predicted failures, and
a bias distribution is fit to these locations (more on this in Section \ref{ss:is}).
SIIS has been deployed with GP surrogates in a variety of subsequent works
\citep[e.g.,][]{cole2023entropy,renganathan2023camera,booth2024actively,
renganathan2024efficient}, and we view it as our primary competitor.  
Yet SIIS has several crucial drawbacks.  We will delay a thorough discussion
to Section \ref{ss:is}, but the highlights are: (i) SIIS relies too heavily on an
arbitrary bias distribution, (ii) it largely ignores the 
carefully trained surrogate, and (iii) it offers no guidance 
for allocating the budget between surrogate training and IS estimation.

\begin{figure}[!ht]
\centering
\includegraphics[width=\linewidth, trim = 40 18 40 22, clip=TRUE]{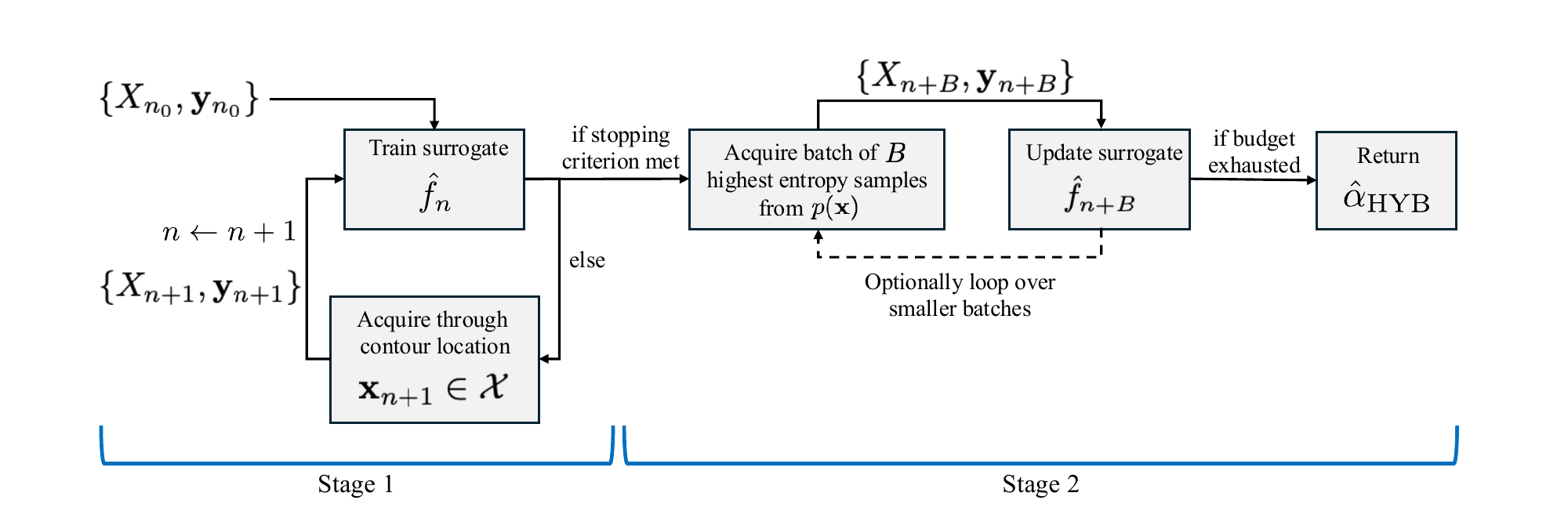}
\caption{Overview of proposed two-stage design for failure probability estimation.
Stage 1 uses contour location to balance exploitation and exploration of 
$\{\mathbf{x}\in\mathcal{X} \mid f(\mathbf{x}) = t\}$.  Stage 2 solely exploits
high uncertainty Monte Carlo samples from $p(\mathbf{x})$. For a fixed budget 
of $n+B$, our approach determines how to effectively allocate $n$ (for contour location) 
and $B$ (for Monte Carlo estimation).}
\label{fig:diagram}
\end{figure} 

We prefer to avoid a bias distribution altogether, instead focusing on how we
could optimally leverage a surrogate and an MC estimator in tandem.  The answer
lies in a two-stage surrogate design, represented in Figure \ref{fig:diagram}
 with details reserved for Section \ref{sec:method}.
First, in Stage 1 we conduct traditional contour location, balancing 
exploration and exploitation while locating the failure contour within domain $\mathcal{X}$.
As training progresses, successive acquisitions will offer less additional insight, indicated
by apparent convergence of the ``surrogate MC'' estimator.  Although we can not be sure 
of theoretical convergence with our limited budget, we propose a stopping criterion to
recognize practical convergence (when surrogate learning from CL has saturated).
Once this stopping criterion is met, instead of continuing contour location or turning to
importance sampling, we propose a novel ``Stage 2'' in which
the remaining simulation budget is spent precisely on the 
{\it Monte Carlo samples from $p(\mathbf{x})$ with the highest classification entropy}.
By observing the true $f(\mathbf{x})$ for samples where our surrogate is uncertain,
we ensure these locations are classified correctly within our final MC estimate.
Ultimately, surrogates trained on the observations acquired in these two unique stages
provide more accurate and precise MC estimates of $\alpha$.
While we prefer to view our contribution through the lens of surrogate design, 
our final failure probability estimate may be considered a ``hybrid MC'' estimator which 
combines the true simulator observations from Stage 2
with surrogate predictions at unobserved samples.

This manuscript is laid out as follows.  Section \ref{sec:review}, 
reviews GP/DGP surrogates, contour location, and importance sampling.
Section \ref{sec:method} presents our two-stage design and failure probability
estimator.  Section \ref{sec:results} presents a variety of synthetic
benchmarks.  Finally, we deploy our method on a computer experiment
of an RAE-2822 airfoil in Section \ref{sec:airfoil} and conclude in Section 
\ref{sec:discuss}.

\section{Preliminaries}\label{sec:review}
  
Denote single
inputs as the row vector $\mathbf{x}$ of size $1\times d$ with response 
$y = f(\mathbf{x})$, where $f$ represents the black-box computer simulation.  
Let $X_n$ represent the row-combined matrix of 
$n$-many inputs with corresponding response vector $\mathbf{y}_n$.  
Our objective is to quantify the 
failure probability from Eq.~(\ref{eq:problem}), namely 
$\mathrm{P}(f(\mathbf{x}) > t)$ for $\mathbf{x}\sim p(\mathbf{x})$. 
Brute force Monte Carlo with a budget of size $M$ provides the estimator
\begin{equation}\label{eq:mc}
\hat{\alpha}_\textrm{MC} = \frac{1}{M} \sum_{i=1}^M \mathbbm{1}_{\left\{f(\mathbf{x}_i)>t\right\}}
    \quad\textrm{for}\quad \mathbf{x}_i\overset{\textrm{iid}}{\sim} p(\mathbf{x}),
\end{equation} 
but this estimator is infeasible when $f$ is expensive.
Let $N$ denote the total budget of expensive simulator evaluations allotted.
We will describe our sample efficient method for estimating failure 
probabilities in Section \ref{sec:method}, targeting $N$ in the hundreds, 
after a quick review of preliminaries.  

\subsection{Gaussian process surrogates}\label{ss:gp}

A surrogate model $\hat{f}$, trained on a limited budget of simulations, may provide
predictions for unobserved inputs in place of the true $f$.  
To enable our proposed methodology, the surrogate must also provide 
uncertainty quantification through posterior predictive distributions.  Gaussian 
processes (GPs) are the predominant choice; they assume a multivariate
normal distribution over the response.  For any $n$ observed locations,
our GP prior assumption is $\mathbf{y}_n \sim\mathcal{N}_n(\boldsymbol\mu, \Sigma(X_n))$.  The 
prior mean $\boldsymbol\mu$ is often simplified to the zero vector after centering (which we will use
moving forward), and the prior covariance $\Sigma(X_n)$ is typically a function
of Euclidean distances, e.g.,
$\Sigma(X_n)^{ij} = \Sigma(\mathbf{x}_i, \mathbf{x}_j) = k(||\mathbf{x}_i - \mathbf{x}_j||^2)$
with kernel $k(\cdot)$.
See \citet{santner2003design,rasmussen2006gaussian,gramacy2020surrogates} for 
thorough reviews.  Conditioned on training data $\left\{X_n, \mathbf{y}_n\right\}$, 
posterior predictions at singleton input $\mathbf{x}$ follow
\begin{equation}\label{eq:gppred}
\hat{f}_n(\mathbf{x}) \sim \mathcal{N}_1\left(\mu_n(\mathbf{x}), \sigma_n^2(\mathbf{x})\right)
\quad\textrm{for}\quad
\begin{array}{rl}
    \mu_n(\mathbf{x}) &= \Sigma(\mathbf{x}, X_n)\Sigma(X_n)^{-1}\mathbf{y}_n \\
    \sigma_n^2(\mathbf{x}) &= \Sigma(\mathbf{x}, \mathbf{x}) - \Sigma(\mathbf{x}, X_n)
        \Sigma(X_n)^{-1}\Sigma(\mathbf{x}, X_n)^\top,
\end{array}
\end{equation}
where $\Sigma(\mathbf{x}, X_n)$ denotes the vector formed from applying the kernel
between $\mathbf{x}$ and every row of $X_n$.
Joint predictions are straightforward but not necessary for our contribution. 

A natural alternative to brute force MC (\ref{eq:mc}) is ``surrogate MC'' which leverages
surrogate predictions in place of the true simulator,
\begin{equation}\label{eq:alphasurr}
\hat{\alpha}_\textrm{SURR}
    = \frac{1}{M} \sum_{i=1}^M \mathbbm{1}_{\left\{\mu_n(\mathbf{x}_i)>t\right\}}
    \quad \textrm{for} \quad \mathbf{x}_i\overset{\textrm{iid}}{\sim} p(\mathbf{x}).
\end{equation}
Here we use the posterior mean of our surrogate $\hat{f}$ in place of the true $f$.
This estimator, however, inherits the bias of the surrogate -- if the surrogate is 
not accurate near the contour, it will misclassify 
inputs which will severely hinder the effectiveness of this estimator.  

\begin{figure}[!ht]
\centering
\includegraphics[width=\linewidth]{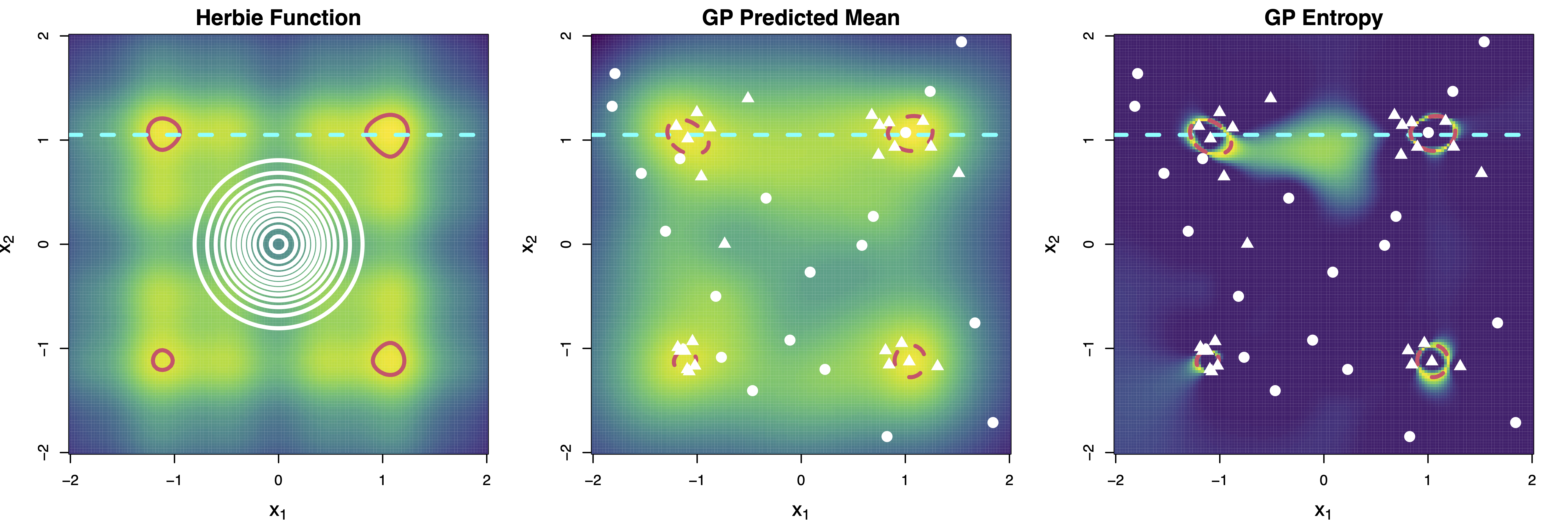}
\caption{{\it Left:} Herbie function (yellow/high, purple/low) with 
true failure contour (red) and input distribution (white). {\it Center/Right:} 
Posterior predicted mean and entropy of a GP trained on an initial 20-point LHS 
(white circles) followed by 30 contour locating acquisitions (white triangles).  
Predicted contour in dashed red.  Dashed line at $x_2=1.05$ indicates
the slice shown in Figure \ref{fig:herbie_slice}.}
\label{fig:herbie}
\end{figure} 

\begin{figure}[!ht]
\centering
\includegraphics[align=t, width=9cm, trim=0 30 0 0]{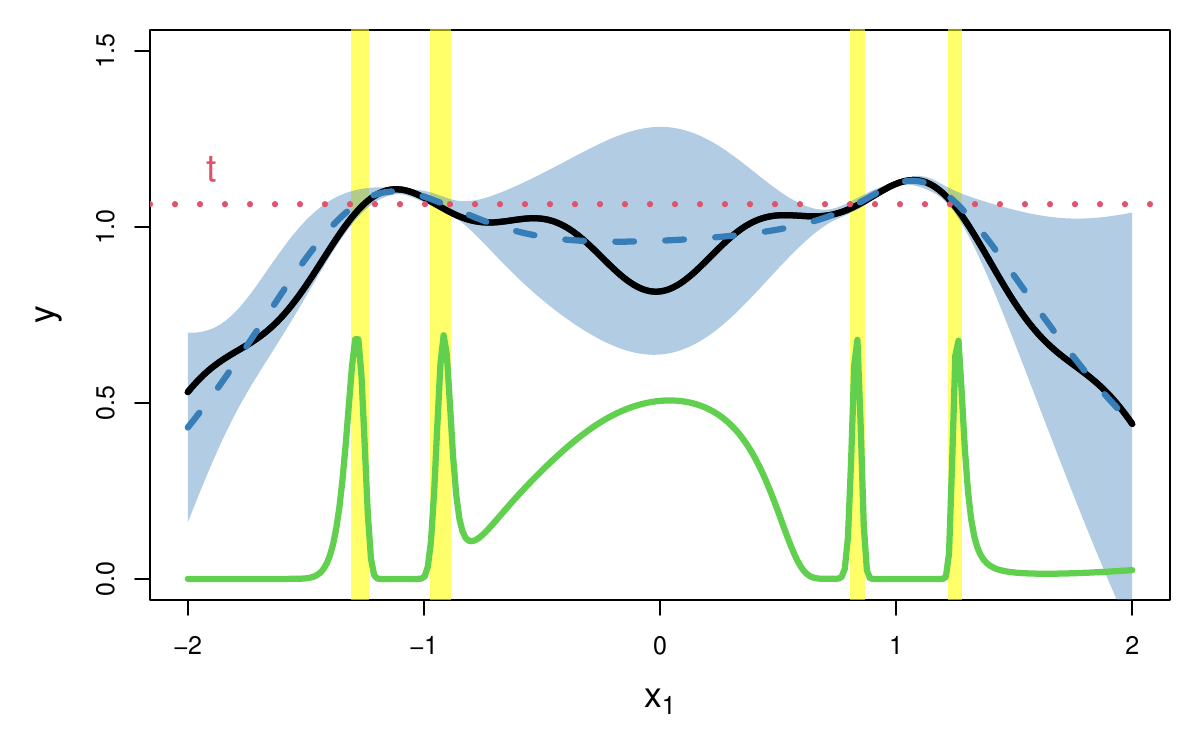}
\includegraphics[align=t, width=5cm, trim=-20 0 0 0]{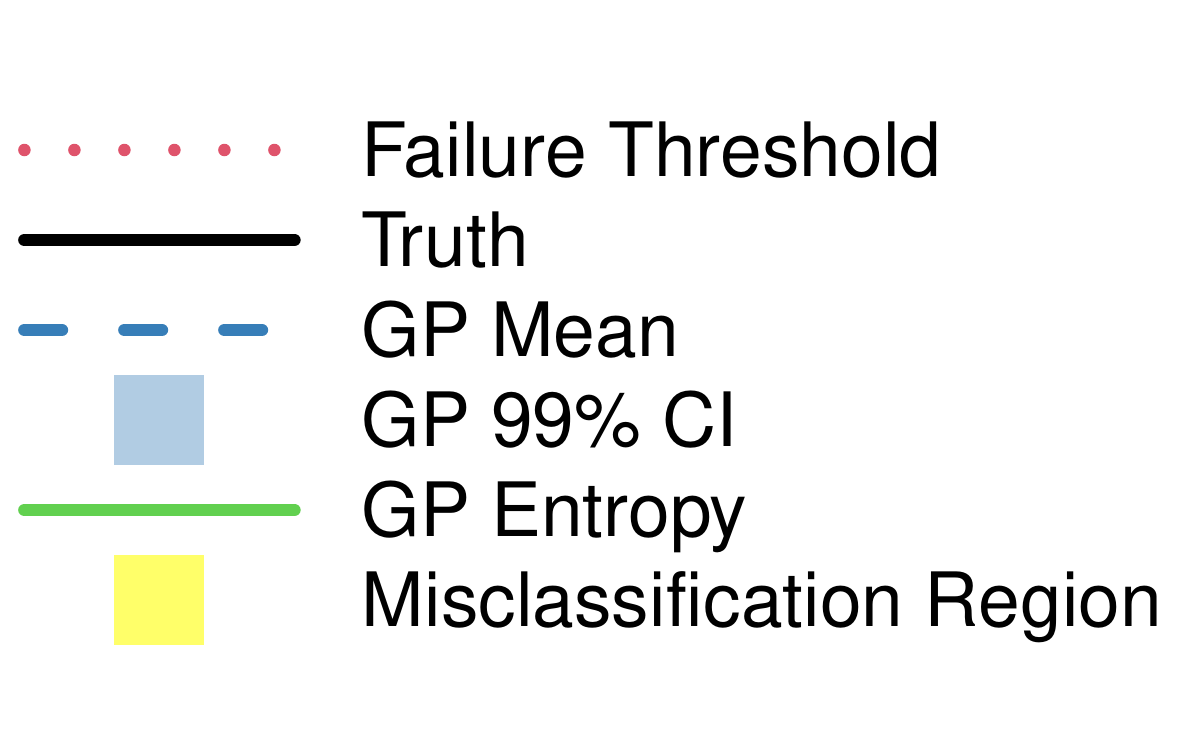}
\caption{Slice of Herbie function along $x_2 = 1.05$ (solid black) with failure threshold 
(dotted red).  GP surrogate from Figure \ref{fig:herbie} shown in blue, with entropy
in green along the $x$-axis.  Yellow shaded regions indicate inputs which the surrogate 
would misclassify.}
\label{fig:herbie_slice}
\end{figure} 

For example, consider the two-dimensional 
``Herbie'' function \citep{lee2011optimization} shown in the left 
panel of Figure \ref{fig:herbie} (we will revisit the other panels
momentarily).  We define a failure threshold of 
$t=1.065$ which demarcates four disjoint failure regions (red contours).  
We also define $p(\mathbf{x})$ using independent normal 
distributions centered at the origin (white contours).\footnote{Full
details of all simulated functions, thresholds, and distributions are 
provided in the Supplementary Material.}  
We trained a GP surrogate using an initial random Latin hypercube 
sample \cite[LHS;][]{mckay2000comparison} of size $n_0=20$ (white circles),
followed by 30 contour locating acquisitions (white triangles, discussed in
Section \ref{sec:cl}).  Figure \ref{fig:herbie_slice} shows the GP posterior
along the slice $x_2 = 1.05$.  The posterior mean (dashed blue) and 
99\% credible interval (blue shading) capture the true nonlinear surface 
(solid black) fairly well.  Even so, there are inputs (highlighted by the
yellow shading) which the surrogate would misclassify as ``passes'' 
instead of ``failures'' or vice versa.  While the predictions in these
regions are reasonably accurate, their proximity to the contour makes precise
classification difficult.

\paragraph{Deep Gaussian processes.} In some situations, surrogate predictions may be
improved through adaptations to the surrogate model itself.  Traditional GPs are limited 
by the commonly assumed stationarity of the 
covariance kernel; they are not able to accurately model nonstationary surfaces
characterized by regime shifts and stark transitions.  Deep Gaussian processes 
\citep{damianou2013deep} are a promising more flexible alternative.  They leverage
functional compositions of GP layers in which latent layers act as warped versions
of the original inputs, allowing for greater flexibility.  DGPs have 
consistently outperformed stationary GPs in modeling nonstationary computer experiments 
\citep[e.g.,][]{rajaram2021empirical,yazdi2022fast,ming2023deep}, particularly when 
deployed in a Bayesian framework using MCMC sampling to 
achieve full posterior integration of latent layers
\citep[e.g.,][]{sauer2023active,sauer2023vecchia}.  See \citet{sauer2023deep} for a 
thorough review of DGP surrogates.  Crucially, our method is agnostic
to surrogate choice as long as UQ is provided.  We will demonstrate functionality
with both traditional GP and Bayesian DGP surrogates in Section \ref{sec:results}.

\subsection{Contour location}\label{sec:cl}
To facilitate failure probability estimation (\ref{eq:alphasurr}), 
a surrogate must accurately classify passes and failures.  Strategic sequential designs 
targeting the failure contour, i.e., ``contour location,'' will outperform 
space-filling counterparts in this regard.  In CL, after initializing with a small
space-filling design, subsequent training locations are selected iteratively through the optimization 
of an acquisition criterion.  The computer simulation is evaluated at the selected
input(s), the surrogate is updated, and the process is repeated until the budget is exhausted
or a stopping criterion is met.  Stage 1 of Figure \ref{fig:diagram}
visualizes this iterative loop.  While acquisitions may be made in batches, we only
consider single acquisitions here.

Identifying acquisition criteria which effectively target failure contours is a hot topic, 
particularly in conjunction with GP surrogates,
\cite[e.g.,][]{bichon2008efficient,ranjan2008sequential,bect2012sequential,chevalier2014fast,
marques2018contour,cole2023entropy}.  Here, we highlight one in particular -- classification 
entropy -- defined as
\begin{equation}\label{eq:ent}
H(\mathbf{x}) = -p_\mathbf{x}\log(p_\mathbf{x}) - (1-p_\mathbf{x})\log(1-p_\mathbf{x}),
\end{equation}
where $p_\mathbf{x}$ represents the predicted probability of failure at input $\mathbf{x}$,
i.e., $\mathrm{P}\left(\hat{f}_n(\mathbf{x}) > t\right)$. 
With a GP surrogate, this predicted probability boils down to a Gaussian CDF computation,
\[
p_\mathbf{x} = \mathrm{P}\left(\hat{f}_n(\mathbf{x}) > t\right) = 
1 - \Phi\left(\frac{t-\mu_n(\mathbf{x})}{\sigma_n(\mathbf{x})}\right),
\]
with $\mu_n(\mathbf{x})$ and $\sigma_n(\mathbf{x})$ following Eq.~(\ref{eq:gppred}).
Revisiting the Herbie example of Section \ref{ss:gp}, we offer two visuals
of the entropy surface for the trained GP.  First, the right panel of Figure \ref{fig:herbie}
shows entropy over the entire 2d space.  Notice high entropy regions are 
predominantly near the failure contour and/or away from training data locations.
Second, Figure \ref{fig:herbie_slice} shows the entropy surface (green)
along the slice $x_2=1.05$.  Entropy is high where the credible interval 
(blue shading) captures the failure threshold (dotted red).  It is influenced
by how much of the credible interval is split above/below $t$, regardless of 
the interval's width. For the highest $x_1$ values, 
surrogate uncertainty is high but entropy is low because the predicted values 
are far from the failure threshold.  As an acquisition criterion, entropy
effectively identifies regions where the surrogate is uncertain of its pass/fail
prediction.  A reasonable acquisition could fall in any of the local optima of the
entropy surface.

Naturally, implementations of entropy as an acquisition function vary on the finer details.
For GP surrogates we will leverage the optimization scheme of \citet{cole2023entropy},
which first deploys a small LHS, identifies the LHS sample with the highest entropy, then
runs a numerical optimization of entropy initialized at that point.  By avoiding multi-starts,
\citeauthor{cole2023entropy} target local optima in the entropy surface.  For DGP
surrogates, we will follow \citet{booth2024contour}.  To avoid cumbersome
numerical optimizations with the heftier MCMC-based DGP, \citeauthor{booth2024contour} use triangulation
candidates \citep{gramacy2022triangulation}, then select the candidate on the Pareto front
of entropy and uncertainty.  Henceforth, all GPs and DGPs 
we consider will be trained with these two contour location methods, respectively.

The right and center panels of Figure \ref{fig:herbie} showed a contour locating 
design for the Herbie function.  Starting with a GP surrogate trained on an LHS 
design of size $n_0=20$ (white circles), we used the entropy-based scheme of \cite{cole2023entropy} 
to acquire 30 more inputs, one-at-a-time (white triangles).
The acquired points attempt to balance exploring the space and exploiting learning
of the failure contour.

\subsection{Importance sampling}\label{ss:is}

Rather than side-stepping the prohibitive computation of Eq.~(\ref{eq:problem}) with
surrogate evaluations, importance sampling \citep{tokdar2010importance}
avoids expensive evaluations by sampling from a different distribution altogether.
Let $q(\mathbf{x})$ denote a ``bias'' density which shares domain with $p(\mathbf{x})$.
The failure probability $\alpha$ may be equivalently represented as
\[
\alpha = \int_{\mathbf{x}\in\mathcal{X}} \mathbbm{1}_{\left\{f(\mathbf{x})>t\right\}} \;
    \frac{p(\mathbf{x})}{q(\mathbf{x})} \;q(\mathbf{x})\, d\mathbf{x}.
\]
Monte Carlo approximation of this integral provides the estimator,
\begin{equation}\label{eq:is}
\hat{\alpha}_\mathrm{IS} = \frac{1}{B}\sum_{i=1}^B w_i \;\mathbbm{1}_{\left\{f(\mathbf{x}_i)>t\right\}}
    \quad\textrm{where}\quad w_i = \frac{p(\mathbf{x}_i)}{q(\mathbf{x}_i)} 
    \quad\textrm{for}\quad \mathbf{x}_i \overset{\textrm{iid}}{\sim}q(\mathbf{x}).
\end{equation}
This estimator requires $B$-many evaluations of the expensive simulator $f$.
The ``bias weights'' $w_i$ account for the difference between the
input density and the bias density.  This estimator is theoretically 
unbiased for any $B > 0$ and requires a smaller budget
of simulator evaluations ($B\ll M$) if the bias distribution appropriately
covers and targets the failure region \citep{srinivasan2002importance}.
But estimation of an effective bias distribution is tricky.
The ideal bias distribution is 
$q(\mathbf{x})^* = \frac{1}{\alpha}\mathbbm{1}_{\left\{f(\mathbf{x})>t\right\}}p(\mathbf{x})$,
but neither $\alpha$ nor the indicator function are known {\it a priori}.
For the sake of feasibility, $q(\mathbf{x})$ is
traditionally constrained to a known distributional family, such as a 
Gaussian mixture model \citep[GMM;][]{reynolds2009gaussian}.  Bias distribution
training is then relegated to learning distributional
parameters (in a GMM, these include the number of mixtures, mixture weights, and 
component means and variances), often accomplished through
cross validation or information criteria maximization.

In ``surrogate-informed importance sampling,'' \citet{peherstorfer2016multifidelity} 
leverage a surrogate to identify
predicted failures from a large Monte Carlo sample of the input distribution, then
train a GMM bias distribution to these predicted failures.  
Recent works have leveraged contour location to train surrogates for this very purpose
\cite[e.g.,][]{cole2023entropy,booth2024actively}.  If the surrogate
effectively identifies the failure region, then these samples should mimic
the ideal $q(\mathbf{x})^*$.  With a trained bias distribution 
in-hand, samples may be drawn from $q(\mathbf{x})$, evaluated through the expensive
simulator, and deployed in Eq.~(\ref{eq:is}) to provide an estimate of the failure probability.
SIIS is a two-stage enterprise (first training a surrogate then
using the surrogate to estimate a bias distribution for importance sampling), but it
is not a ``two-stage design'' since the final estimator makes no use of the surrogate
nor the observations that had been used to train the surrogate.

Although SIIS has been widely embraced for its unbiasedness, it has several key weaknesses.  
First, the bias distribution, which features heavily in the final estimate of Eq.~(\ref{eq:is}), 
is a completely auxiliary quantity.  Hypothetically, two different bias
distributions could produce the same samples 
$\left\{\mathbf{x}_i\right\}_{i=1}^B$, thus producing the same
$\left\{f(\mathbf{x}_i)\right\}_{i=1}^B$.  But in the calculation of Eq.~(\ref{eq:is}), 
these two bias distributions would contribute different weights, resulting in different
final answers even though the training data from the computer simulator was exactly
the same.  \citet{ohagan1987monte} first pointed out this conundrum as a 
violation of the ``Likelihood principle.''  A second weakness is the restriction
of $q(\mathbf{x})$ to distributional families like GMMs.  Failure regions may be complex
and may not fit nicely under these target distributions.  Consequently, SIIS typically 
requires large $B$ in order to adequately cover the failure region.  
For context, \citet{peherstorfer2016multifidelity}
entertained $B \in\left\{100, 1000, 10000, 100000\right\}$ and found performance continued to
improve as $B$ increased.  Furthermore, there is no clear avenue for determining
the budget allocation between surrogate training and importance sampling estimation.
[In our benchmark exercises, we will borrow from our proposed method to inform 
budget allocation for the SIIS competitor.]
Finally, the SIIS method does not leverage the full potential
of the surrogate model.  Many expensive simulator evaluations go into training 
the surrogate, but the surrogate is abandoned after it has identified
potential bias locations for estimation of $q(\mathbf{x})$.

\section{Two-stage design for failure probability estimation}\label{sec:method}

\begin{figure}[!ht]
\centering
\includegraphics[width=0.85\linewidth]{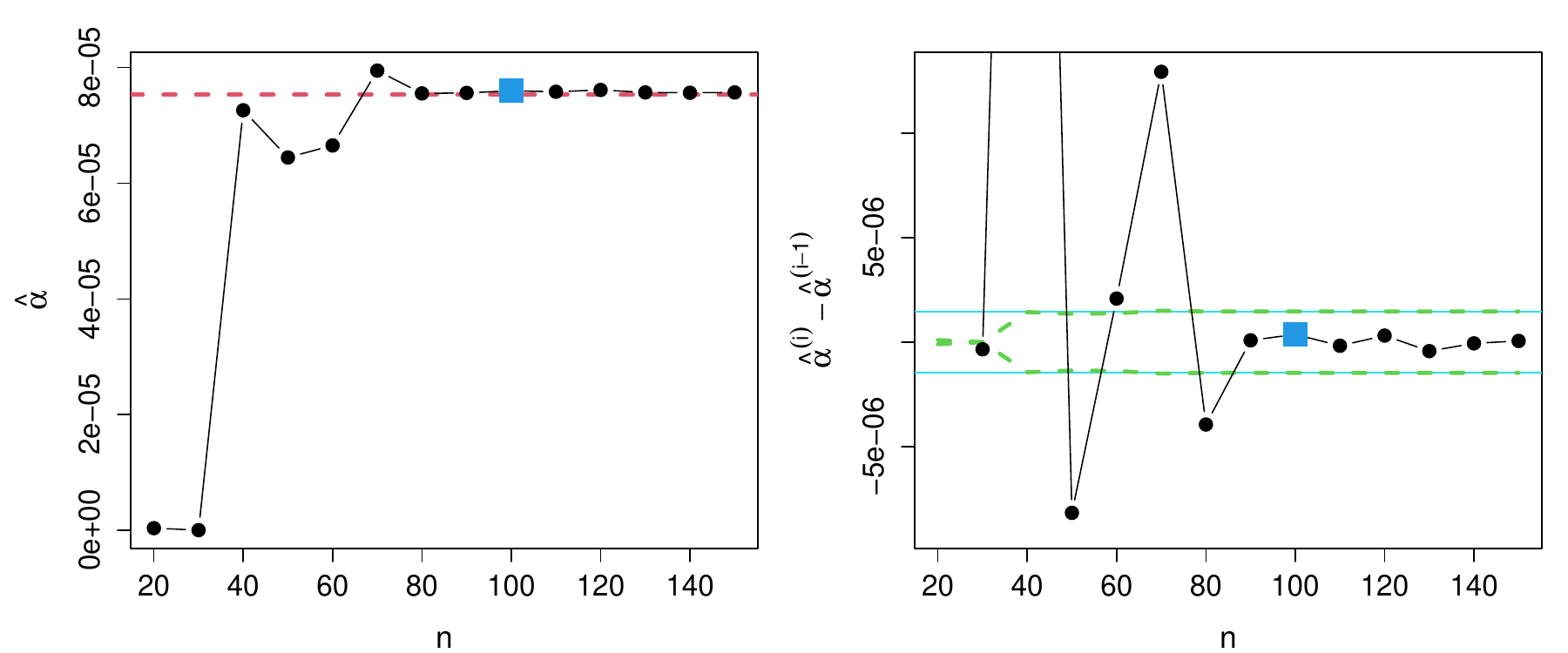}
\caption{{\it Left:} Failure probability estimate (\ref{eq:alphasurr}) from
a GP surrogate of the Herbie function (as shown in Figure \ref{fig:herbie}), 
updated after every 10th acquisition of the CL design.  True $\alpha$ in
dashed red. {\it Right:} Difference in successive $\hat{\alpha}$ estimates with 
$\pm\hat{\sigma}_\alpha$ shown in dashed green.  Narrow teal line indicates true $\sigma_\alpha$.
Blue square marks the second update within $\pm\hat{\sigma}_\alpha$, 
which satisfies our proposed stopping criterion.}
\label{fig:progress}
\end{figure}

Ultimately, we are unsatisfied with the two aforementioned surrogate-based methods for 
estimating $\alpha$: exhaustive CL with surrogate MC and 
surrogate-informed importance sampling.  
While the drawbacks to SIIS are apparent, the limitations of exhaustive CL
deserve more attention.
To investigate further, we revisit the CL design of the Herbie
function (Figure \ref{fig:herbie}), this time continuing contour location
until an entire budget of $N=150$ has been exhausted.  As the design progresses, 
we evaluate and store the surrogate MC estimate $\hat{\alpha}_\textrm{SURR}$ 
(\ref{eq:alphasurr}), which we now denote as simply $\hat{\alpha}$, after every 
10th CL acquisition.  Progress in this estimate is shown in the left panel of 
Figure \ref{fig:progress}.  While there is significant learning occurring
with early acquisitions, it quickly plateaus.  By the time the surrogate
has been trained on 100 points, it knows just as much about $\alpha$ as it
does when it has been trained on the full budget of $N=150$ points.  
In this case, continued contour location is effectively wasting the 
remaining simulation budget.
We propose a solution: first developing a stopping criterion to decide when to 
halt contour location (Figure \ref{fig:diagram}, Stage 1), then proposing a 
second design stage (Figure \ref{fig:diagram}, Stage 2) 
which greedily leverages the remaining simulation budget for Monte Carlo
samples with the highest classification uncertainty.  The second stage,
in which observations are acquired directly from $p(\mathbf{x})$
enables a``hybrid Monte Carlo'' estimator, using both true simulator observations
and surrogate predictions.

\subsection{A stopping criterion for contour location}\label{sec:stop} 

Let $n$ denote the budget of expensive simulator evaluations used on contour 
location (Stage 1).  Let $B$ denote the remaining evaluation budget,
such that the total budget is $N=n+B$.
We seek an objective, automatable decision rule for selecting $n$, halting CL
once surrogate learning has plateaued.  Consider the progress in $\hat{\alpha}$ 
shown in Figure \ref{fig:progress}.  The right panel shows the
relative difference between each successive $\hat{\alpha}$
estimate from the left panel.  As expected, the relative differences converge 
to zero as the surrogate hones in on the true contour, but with this finite budget
they never settle at exactly zero.
We need an objective method to decide when these updates to $\hat{\alpha}$ 
are no longer significant.

The challenge is that $\hat{\alpha}$ exhibits variability
from two distinct sources: surrogate error and the natural variability that accompanies
any Monte Carlo estimator with finite $M$.  Hypothetically, presume we could
evaluate the true $f$ an unlimited number of times to obtain the brute force MC
estimator of Eq.~(\ref{eq:mc}).  In this estimator, 
the indicator $\mathbbm{1}_{\left\{f(\mathbf{x})>t\right\}}$ is a Bernoulli 
random variable with expected value $\alpha$ and variance $\alpha(1-\alpha)$.  
Thus, the MC estimator has variance inversely proportional to $M$:
\[
\mathbb{V}[\hat{\alpha}_\textrm{MC}] = \frac{1}{M^2}\sum_{i=1}^M
    \mathbb{V}\left[\mathbbm{1}_{\left\{f(\mathbf{x})>t\right\}}\right] = 
    \frac{\alpha(1-\alpha)}{M}.
\]
Let $\sigma_\alpha = \sqrt{\frac{\alpha(1-\alpha)}{M}}$ denote the standard
error of this estimator; it quantifies the degree of variability that we would
expect even if we used the true $f$ (i.e., zero surrogate error).
We consider successive updates to $\hat{\alpha}$ within this realm of intrinsic MC 
variability to indicate stabilization of the surrogate.  Since the true 
$\alpha$ is unknown, we estimate 
$\hat{\sigma}_\alpha = \sqrt{\frac{\hat{\alpha}(1-\hat{\alpha})}{M}}$
using the current surrogate MC estimate.  While this does intertwine surrogate
error and MC error, it still provides a useful and feasible estimate.

To demonstrate, the green dashed lines in the right panel of Figure 
\ref{fig:progress} show $\pm\hat{\sigma}_\alpha$, updated with $\hat{\alpha}$
after every 10th acquisition.
The estimated $\hat{\sigma}_\alpha$ (dashed green) effectively reflects the 
true $\sigma_\alpha$ (narrow solid teal)
after very few acquisitions.  At the start of the design, updates to $\hat{\alpha}$ 
are far beyond this estimated standard error.  But after a while, the 
updates sit comfortably within these bounds.  

Let $X_M$ denote the row-combined matrix of $M$-many samples 
from $p(\mathbf{x})$.  We posit that, for fixed $X_M$, any update to $\hat{\alpha}$ 
within $\pm\hat{\sigma}$ is insignificant, as it is within the realm of expected 
MC error.  [To avoid excessive computation, we recommend updating estimates after 
every 10 acquisitions, which we will do in all later exercises.] Accordingly, we propose halting 
CL once we have observed two successive updates within $\pm\hat{\sigma}$.
The blue square in Figure \ref{fig:progress} marks this point, stopping CL for this
example at $n=100$, effectively identifying a point where surrogate learning
has saturated and leaving a remaining budget of $B=50$.
Note, this stopping criterion requires the same $X_M$ be used in every
$\hat{\alpha}$ calculation.  If a new $X_M$ is generated for 
each estimate, there would be additional sampling variability reflected
among the $\hat{\alpha}$'s (on top of the surrogate error and MC variability).

Our budgeting scheme is summarized as follows.  Initialize a surrogate,
begin a contour locating sequential design, and collect one sample $X_M$.  
At the start, it is likely that
there will be no failures observed and that $\hat{\alpha}$ will be equal to 
zero (as seen in Figure \ref{fig:progress}).  To safeguard 
against halting surrogate training prematurely, we build
in a ``common sense'' check by setting a minimum number of failures that must
be observed before contour location could be halted.
If the contour locating design is succeeding, it will be placing some points in the
failure region.  We recommend a default of 10 failures, but suggest increasing
this for more complex, higher dimensional problems.
Once the design has surpassed these minimum requirements, estimate 
$\hat{\alpha}$ after every 10th acquisition.  Calculate $\hat{\sigma}_\alpha$ 
and check if  
$|\hat{\alpha}^\textrm{(current)} - \hat{\alpha}^\textrm{(previous)}| < \hat{\sigma}_\alpha$.
When this condition has been satisfied twice in a row, halt the CL
design of Stage 1 and proceed to Stage 2 (the topic of Section \ref{sec:hybmc}).

\paragraph{Choosing $M$.}  In practice, it is important to select a MC sample size
that is large enough without being too cumbersome.
Larger $M$ will produce more effective sum approximations of 
the true integrals, but we need surrogate predictions (e.g., Eq.~\ref{eq:gppred}) for 
every sample in $X_M$.  Although this computation is significantly cheaper than 
evaluation of the expensive simulator, it could be significant for extremely large $M$, 
particularly with a heftier surrogate like a Bayesian DGP.

If we knew $\alpha$, we could determine $M$ by
thresholding the standard error $\sigma_\alpha$.  For example, to achieve 
a standard error less than 10\% of $\alpha$ requires:
\[
\sigma_\alpha < \frac{\alpha}{10}
\quad\longrightarrow\quad \frac{100(1-\alpha)}{\alpha} < M.
\]
To achieve this precision, $\alpha=10^{-4}$ requires $M$ nearly 1 million,
$\alpha=10^{-5}$ requires $M$ nearly 10 million, and 
$\alpha=10^{-6}$ requires $M$ nearly 100 million.
In practice, with unknown $\alpha$, we recommend either choosing
$M$ conservatively given expert insight into the anticipated failure probability 
or performing a pilot study where $M$ is incremented say by one million at a time.
Gather surrogate predictions along this incremental process, and use
$\hat{\alpha}$ to inform $\hat{\sigma}_\alpha$.

\subsection{Stage 2: Hybrid Monte Carlo for failure probability estimation}\label{sec:hybmc}

Once $n$ is determined and the contour location of Stage 1 is complete, how should we spend the 
remaining budget of simulator evaluations ($B=N-n$) in Stage 2?
Recall, our objective is to leverage the trained surrogate to estimate $\alpha$ (\ref{eq:problem}).
We propose a new design for Stage 2, which enables an upgrade to the surrogate MC estimator of 
Eq.~(\ref{eq:alphasurr}).

To set the stage, let $\mathcal{U}\subset\mathcal{X}$ represent some subset of the input domain (boundary inclusive).  
Then, Eq.~(\ref{eq:problem}) may be partitioned as
\[
\alpha=\int_\mathcal{U} \mathbbm{1}_{\{f(\mathbf{x})>t\}}p(\mathbf{x})\, d\mathbf{x} +
\int_\mathcal{\mathcal{X}\setminus \mathcal{U}} \mathbbm{1}_{\{f(\mathbf{x})>t\}}p(\mathbf{x})\, d\mathbf{x},
\]
inspiring an equivalent formulation of the surrogate MC estimator,
\[
\hat{\alpha}_\textrm{SURR} = 
    \frac{1}{M}\left[\sum_{\mathbf{x}_i\in\mathcal{U}} 
    \mathbbm{1}_{\left\{\mu_n(\mathbf{x}_i) > t\right\}} + \sum_{\mathbf{x}_i\notin\mathcal{U}} 
    \mathbbm{1}_{\left\{\mu_n(\mathbf{x}_i) > t\right\}}\right]
    \quad\textrm{for}\quad \mathbf{x}_i\overset{\textrm{iid}}{\sim} p(\mathbf{x}).
\]
Now for $\mathbf{x}_i\in\mathcal{U}$, we upgrade surrogate prediction $\mu_n(\mathbf{x}_i)$ 
to expensive evaluation of the true $f(\mathbf{x}_i)$,
resulting in the following ``hybrid MC'' estimator:
\begin{equation}\label{eq:hybmc}
\hat{\alpha}_\textrm{HYB} = \frac{1}{M}\left[\sum_{\mathbf{x}_i\in\mathcal{U}} 
    \mathbbm{1}_{\left\{f(\mathbf{x}_i) > t\right\}} + \sum_{\mathbf{x}_i\notin\mathcal{U}} 
    \mathbbm{1}_{\left\{\mu_n(\mathbf{x}_i) > t\right\}}\right]
    \quad\textrm{for}\quad \mathbf{x}_i\overset{\textrm{iid}}{\sim} p(\mathbf{x}).
\end{equation}
The hybrid nature of this estimator is in the use of both true 
simulator evaluations (left term) and surrogate evaluations (right term).  
The accuracy of this estimator is contingent upon the
classification accuracy of the surrogate and the choice of $\mathcal{U}$.
If we restrict the size of $\mathcal{U}$ so only $B$-many samples from $X_M$ fall
within it, i.e., $\sum_{i=1}^M \mathbbm{1}_{\left\{\mathbf{x}_i\in\mathcal{U}\right\}}=B$,
then we can use this hybrid estimator while staying within our budgetary constraints.
The evaluation of the expensive simulator at the Monte Carlo samples within $\mathcal{U}$ 
constitutes the second stage of our design.

Rather than thinking of $\mathcal{U}$ as a continuum, we find it easier to 
conceptualize in the context of the discrete set $X_M$.
In a perfect world, $\mathcal{U}$ would capture all the samples which the 
surrogate would misclassify.  We are not privy to this information, and we are 
under a limited budget, but we can leverage the surrogate's UQ to identify the 
$B$-many samples at which it is most uncertain.
Recall, classification entropy (\ref{eq:ent}) quantifies the degree of uncertainty
in the pass/fail prediction of a surrogate.  Consequently, our proposed allocation of samples
to $\mathcal{U}$ is strikingly simple -- place the $B$-many samples 
from $X_M$ with the highest entropy into $\mathcal{U}$.  Denote these as 
$X_M^B$.  Revisiting Figure \ref{fig:herbie_slice}, notice the high entropy 
regions (green) align well with the misclassification regions 
(yellow shading).  The entropy criterion is effectively identifying regions of 
high classification uncertainty where we could most utilize expensive true 
simulator evaluations to correct inaccurate surrogate classifications. 

Ultimately, after completing the second design stage by
evaluating $f(\mathbf{x})$ for $\mathbf{x}\in X_M^B$, 
we update the surrogate with these additional observations, thereby updating
$\mu_n(\mathbf{x})$, before returning 
$\hat{\alpha}_\textrm{HYB}$ (\ref{eq:hybmc}).  The process of observing the
true simulator at the high entropy locations and updating the surrogate accordingly
could be done in small batches within a feedback loop as demonstrated in 
Figure \ref{fig:diagram}.  Yet in our exercises, we have not seen smaller batch sizes
to improve performance (we provide empirical evidence of this in Supplementary Material).
Accordingly, we prefer to collect a single batch of all $B$-many high entropy samples at once.

\subsection{Summary and related work}  

We have presented a comprehensive strategy for estimating Eq.~(\ref{eq:problem}) from
a budget of size $N$.  It is summarized as follows.  Start with a very large Monte
Carlo sample from $p(\mathbf{x})$, denoted $X_M$.  Train a surrogate model
through a contour locating sequential design, monitoring the progress of the 
$\hat{\alpha}_\textrm{SURR}$ estimate (\ref{eq:alphasurr}) for given $X_M$, and halting
contour location once two consecutive updates are within $\pm\hat{\sigma}_\alpha$.
Denote the design points chosen in this stage as $X_n$.  Then, use the trained
surrogate to identify the highest entropy samples, $X_M^B$, evaluate the simulator
at these inputs, and update the surrogate with these new observations.
Finally, return $\hat{\alpha}_\textrm{HYB}$ following Eq.~(\ref{eq:hybmc}).

Given the deterministic nature of $f(\mathbf{x})$, effective surrogates should
interpolate observed points, i.e., $\mu_n(\mathbf{x}_i)=f(\mathbf{x}_i)$ when the
surrogate has been trained on $\{\mathbf{x}_i,y_i\}$.  In this light, the hybrid MC estimator (\ref{eq:hybmc}) 
is akin to a surrogate MC estimator (\ref{eq:alphasurr}) when the surrogate has been 
trained on all $\mathbf{x}_i\in\mathcal{U}$.  This connection permits us to frame
our contribution as a {\it two-stage design}.  In the first stage, $X_n$
is chosen over $\mathcal{X}$ based on contour location.  
In the second stage, $X_M^B$ is chosen over $\mathbbm{P}_\mathbf{x}$ based on maximum entropy.
The final design of size $N$ is then $X_N = [X_n^\top, {X_M^B}^\top]^\top$.

There are some noteworthy distinctions between these two design stages, as depicted
in Figure \ref{fig:diagram}.
In the first stage, the $n$ points selected for contour location may be acquired 
anywhere in the domain $\mathcal{X}$, seeking to balance both exploration of the
response surface and exploitation of the suspected contour.  In fact, the contour 
locating sequential design strategies we consider in Section \ref{sec:results} 
\citep{cole2023entropy,booth2024contour}
intentionally circumvent acquisitions based on maximal entropy in order to avoid
clustered acquisitions and to promote more exploration of the surface.
Alternatively, the $B$ points chosen in the second stage must originate 
from $p(\mathbf{x})$ and are solely exploitative.  By nature,
they will be clustered around the predicted contour, where pass/fail classifications
are hardest to pin down.  We contend that having a mix 
of data from these two realms is superior to either extreme.

Some of the previous works referenced in Section \ref{sec:intro} have 
investigated stopping criteria for surrogate training in reliability settings.
The most similar to our proposed criterion is that of \citet{zhu2016reliability}
who recommend a convergence criterion akin to
$\frac{\hat{\sigma}_\alpha}{\hat {\alpha}} \leq \epsilon$, 
where $\epsilon$ is a user-specified threshold.  While similar to our 
stopping criterion, the choice of $\epsilon$ here is make-or-break, and little 
guidance is provided.  If $\epsilon$ is 
too high, contour location will be halted too soon; if too low, 
contour location will continue past the point of utility.  
Our proposed stopping criterion avoids this choice and may be 
easily automated without problem-specific tuning.  In our synthetic 
experiments, we find that it consistently results in appropriate 
termination with budget to spare.

Hybrid MC estimators (similar to Eq.~\ref{eq:hybmc}) have been entertained
in other settings. \citet{li2010evaluation} use polynomial chaos models
for a hybrid MC estimator with samples allocated
to $\mathcal{U}$ based on predicted proximity to the contour, i.e., $|\mu_n(\mathbf{x}_i) - t|$.
Their work, however, does not consider surrogate design and does 
not leverage surrogate UQ.
\citet{fuhrlander2020blackbox} use hybrid MC, but they only 
consider training the surrogate on samples from $p(\mathbf{x})$.  
\citet{echard2011ak} similarly train only on a subset of $X_M$.  We attempted
to recreate the approach of \citet{fuhrlander2020blackbox} in our 
setting: (i) sample large $X_M$, (ii) train an initial 
surrogate on a random selection of these samples, 
(iii) use the surrogate with a metric such as entropy to select additional 
batches from $X_M$ for evaluation through the simulator, (iv) update the
surrogate, and (v) repeat steps (iii)-(iv) until the budget is exhausted.
Yet we were unable to obtain even reasonable results because initial surrogate
fits were very poor, and surrogate-informed acquisitions were unable to
recover.  Random selections from $X_M$ were overwhelmingly 
in high probability regions, not low probability failure regions.
\citet{hristov2019adaptive} noticed this conundrum and stated it well:
``using a surrogate that is built on frequent events to predict and
analyze rare events'' can be futile.  

To visualize this conundrum in a simple setting, we revisit the Herbie 
example of Figure \ref{fig:herbie}.  With the input distribution centered at the origin,
the probability of generating any samples outside $[-1, 1]^2$ is very low.  But 
a GP surrogate would benefit from some ``anchor points'' in these outer regions 
\citep{gramacy2015local}.  This distinction will be less prominent if $p(\mathbf{x})$
is uniformly distributed over $\mathcal{X}$, but the trade-off between exploring in 
Stage 1 and exploiting in Stage 2 will still be impactful.
Even with uniform $p(\mathbf{x})$, training a surrogate only on maximal entropy
points is suboptimal \citep{cole2023entropy,booth2024contour}.

\section{Synthetic experiments}\label{sec:results}

In this section we benchmark our method against state-of-the-art competitors
on a variety of synthetic experiments.
We consider four functions: the 2d Herbie function 
\citep{lee2011optimization} with a total budget of 150, the 3d 
Ishigami function \citep{ishigami1990importance} with a total
budget of 300, the 6d Hartmann function \citep{picheny2013benchmark} 
with a total budget of 600, and the 4d Plateau function \citep{booth2024contour}
with a total budget of 200.  The first three of these are plausibly stationary,
so we use a standard GP surrogate, fit with the
{\tt Scikit-learn} package in {\sf python} \citep{pedregosa2011scikit}, 
and the entropy contour location scheme 
of \citet{cole2023entropy}. The Plateau function is known to be nonstationary; it is 
characterized by flat regions with a steep sloping drop between them.  To address 
the nonstationarity, we use a Bayesian DGP surrogate \citep{sauer2023active},
fit with the {\tt deepgp} package in {\sf R} \citep{deepgp},
and the contour location scheme of \citet{booth2024contour}.  With the right surrogate,
the failure contour of the Plateau function can be well estimated 
from a rather small amount of data.

\begin{table}[h!]
\begin{center}
\begin{tabular}{c|c|cccc|ccc}
Function & True $\alpha$ & $d$ & $n_0$ & $N$ & $M$ & Surrogate & CL Scheme \\
\hline
Herbie & $7.533\times 10^{-5}$ & 2 & 20 & 150 & $3.5\times 10^7$ & GP & \citet{cole2023entropy} \\
Ishigami & $1.904\times 10^{-4}$ & 3 & 50 & 300 & $1.5\times 10^7$ & GP & \citet{cole2023entropy}  \\
Hartmann & $1.001\times 10^{-5}$ & 6 & 100 & 600 & $1.0\times 10^8$ & GP & \citet{cole2023entropy} \\
Plateau & $4.308\times 10^{-4}$ & 4 & 30 & 200 & $3.5\times 10^6$ & DGP & \citet{booth2024contour} 
\end{tabular}
\end{center}
\vspace{-5mm}
\caption{Simulation settings. True $\alpha$ values are estimated from 10 billion samples.}
\label{tab:settings}
\end{table}

Table \ref{tab:settings} summarizes the simulation settings.  Functions are observed without noise to 
mimic a deterministic computer simulation.  We define failure thresholds and input
distributions for each function, resulting in the failure probabilities reported 
in Table \ref{tab:settings}.  All input distributions are variations/combinations of independent 
uniform and truncated normal distributions; further details of the 
functions, thresholds, and distributions are reserved for Supplementary Material.
We choose $M$ for each $\alpha$ to keep $\sigma_\alpha$ conservatively under 5\%.

\begin{figure}[!ht]
\centering
\includegraphics[width=\linewidth]{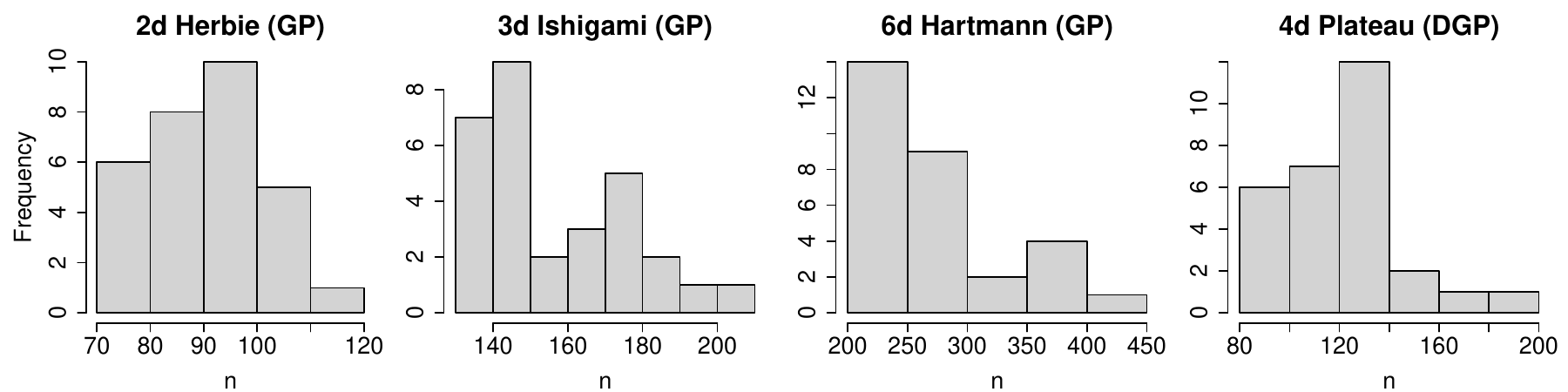}
\caption{Distribution of the chosen $n$ across 30 repetitions.}
\label{fig:budget}
\end{figure} 

First, we deploy our stopping criterion (Section \ref{sec:stop}) for each of
the four scenarios (Table \ref{tab:settings}).  Starting with an initial LHS 
sample of size $n_0$, we proceed with Stage 1, conducting CL with the 
specified method and determining the $n$ at which to stop.
We conduct 30 re-randomized Monte Carlo repetitions. 
To avoid premature termination, we require a minimum of 10 observed 
failures and $n \geq 2n_0$ (at least as many contour locating acquisitions as original
space-filling observations).  Figure \ref{fig:budget} shows the distribution of chosen 
$n$ values for each scenario.  
Notice, all trials terminated before the allotted $N$, reserving some evaluations
for Stage 2, but there was considerable variability across trials.
We view this variability as a strength -- some random initializations need
more data to effectively locate the contour while some get lucky from the start.
Allowing the surrogate to dictate its own budget allocation mitigates some of the risks
of having an unlucky starting design.   

For each trained surrogate with chosen $n$ (which varies across repetitions), we
collect a large Monte Carlo sample ($X_M$, constant across methods, randomized 
across repetitions), then compare the following estimation methods.  
All variations use the same Stage 1 contour location scheme up to point $n$; 
they differ only in their use of the remaining $B=N-n$ observations.
\begin{itemize}
    \item Exhaustive CL: Ignoring the chosen $n$, continue contour location 
    until the entire budget is exhausted and return the surrogate MC estimate 
    (\ref{eq:alphasurr}).
    \item Two-stage Entropy: Our {\bf proposed} two-stage approach 
    (Figure \ref{fig:diagram}) using the hybrid MC estimate 
    (\ref{eq:hybmc}) where $\mathcal{U}$ is chosen to contain the $B$ points from 
    $X_M$ with highest entropy. 
    \item Two-stage Proximity: Same as ``Two-stage Entropy,'' but $\mathcal{U}$
    is chosen to contain the $B$ points from $X_M$ which are predicted to be closest
    to the failure threshold (i.e., the samples with the smallest $|\mu_n(\mathbf{x}_i) - t|$).
    This is inspired by the work of \citet{li2010evaluation}.
    \item SIIS: Surrogate-informed importance sampling (\ref{eq:is}), with $B$-many samples
    observed from an estimated GMM bias distribution \citep{peherstorfer2016multifidelity}.  
    The bias distribution is trained on the predicted failures, i.e.,
    $\left\{\mathbf{x}_i\in X_M \;|\; \mu_n(\mathbf{x}_i) > t\right\}$.  For our GP surrogates
    implemented in {\sf python}, we use the {\tt scikit-learn} package \citep{pedregosa2011scikit} 
    to train GMM bias distributions.  For our DGP surrogates implemented in {\sf R}, we use the 
    {\tt mclust} package \citep{mclust}.
    \item SIIS UCB: Same as ``SIIS,'' but the bias distribution is trained on the samples whose
    upper 95\% confidence bound exceeds the threshold, i.e.,
    $\left\{\mathbf{x}_i\in X_M \;|\; \mu_n(\mathbf{x}_i) + 1.645\sigma_n(\mathbf{x}_i) > t\right\}$.
\end{itemize}
As implemented, the latter three are to some degree an internal comparison since they
use our proposed stopping criterion to determine the budget allocation between surrogate
training ($n$) and estimation ($B$).  Without borrowing from our proposed approach, there would
be no guidance on the allocation of $n$ and $B$ for these methods.  The ``Two-stage Proximity''
approach is truly an internal comparison as it borrows everything from our proposed
two-stage design aside from the specific allocation of samples to $\mathcal{U}$.  It is
included to show the impact of the entropy selection criterion, which uses surrogate UQ,
over a selection solely based on the posterior mean.  Reproducible code for all experiments, 
including both the {\sf python} and {\sf R} implementations, is available in our public git 
repository.\footnote{\url{https://bitbucket.org/boothlab/failprob/}}

\begin{figure}[ht!]
\centering
\includegraphics[width=0.47\linewidth]{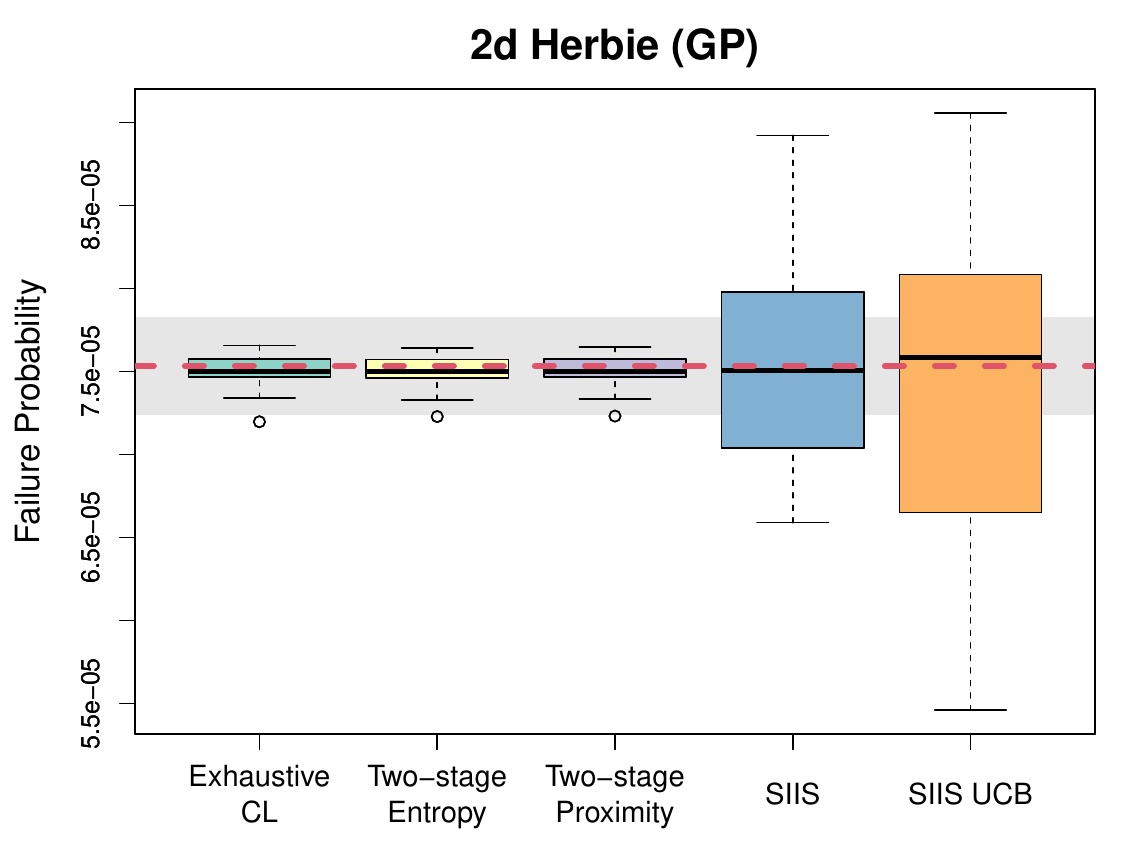}
\includegraphics[width=0.47\linewidth]{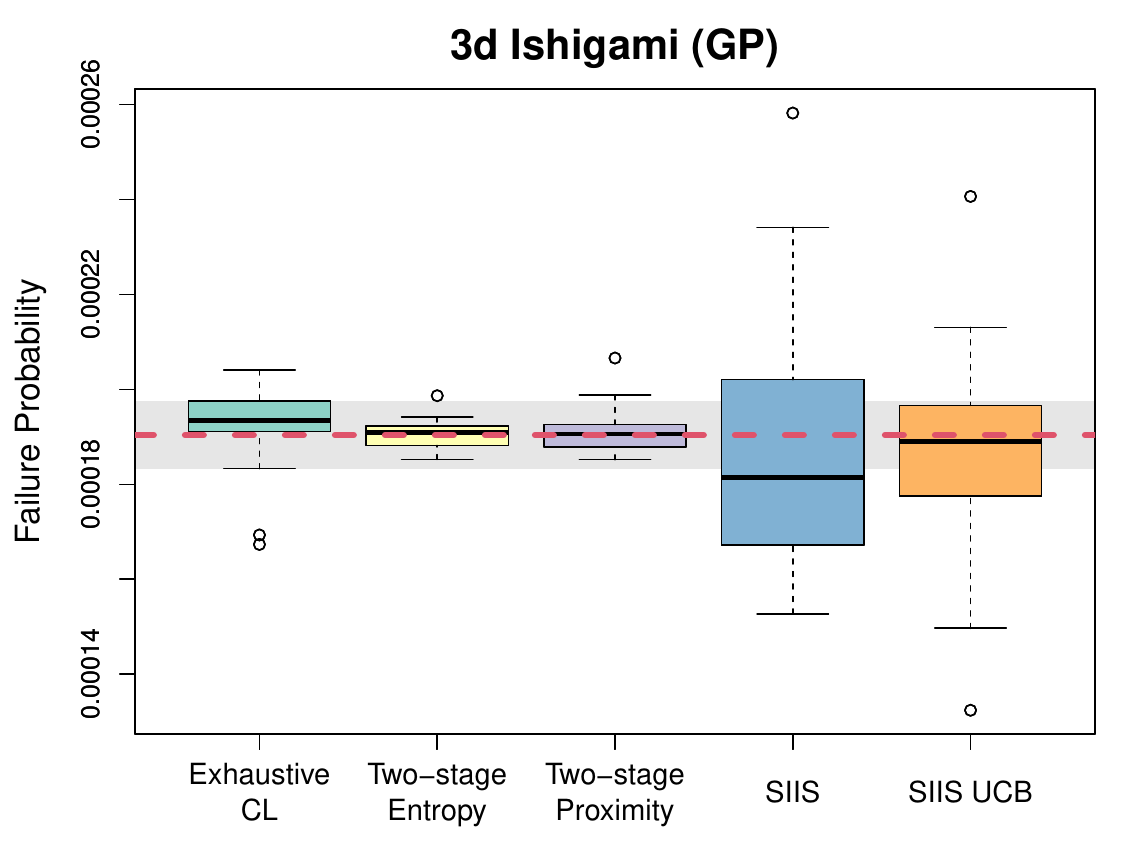}\\
\includegraphics[width=0.47\linewidth]{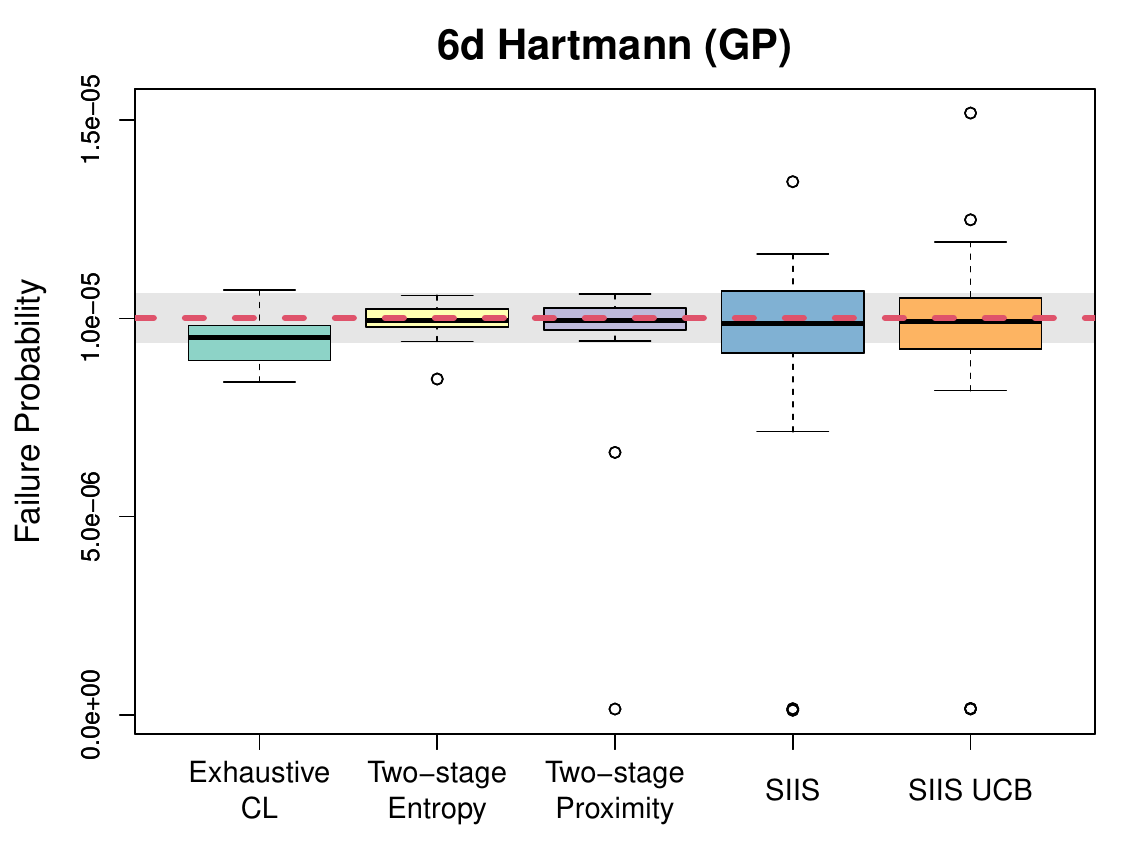}
\includegraphics[width=0.47\linewidth]{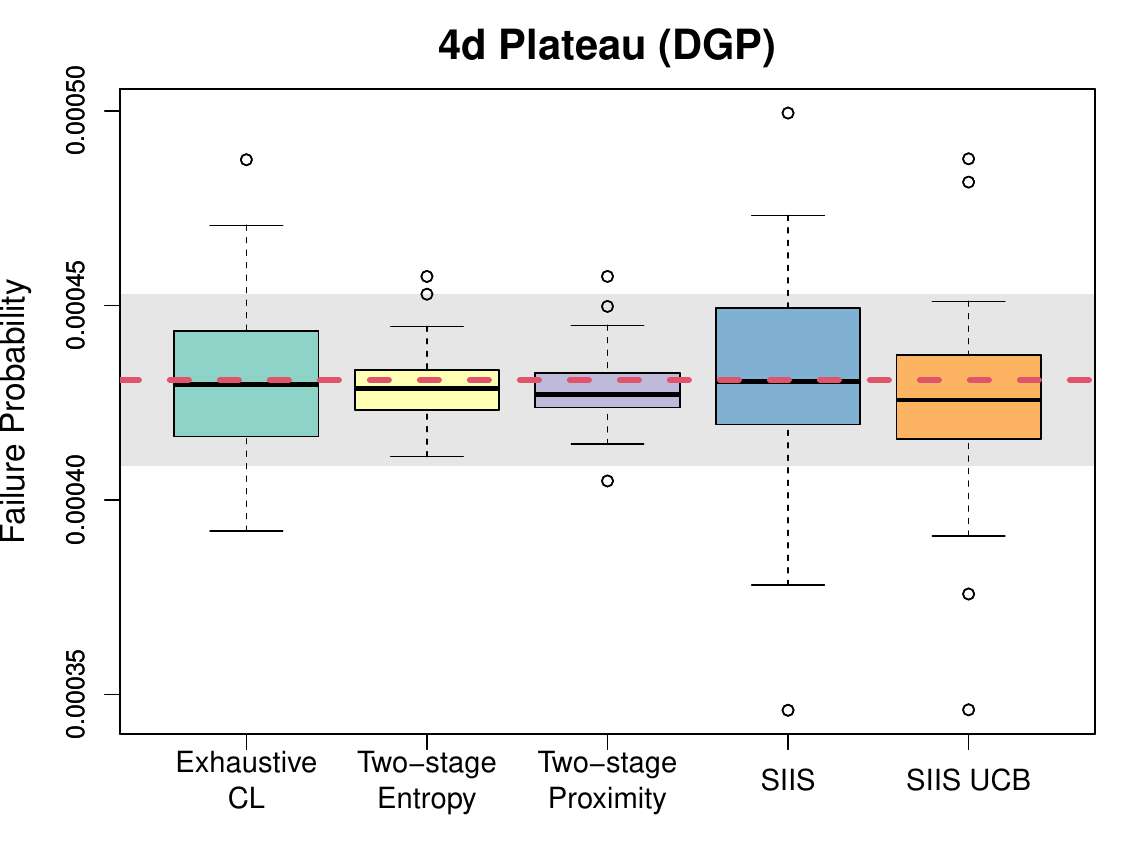}\\
\caption{Estimated failure probabilities shown over true failure probability 
(red dashed) $\pm 2$ standard errors (grey shading).  Boxplots show 30
repetitions.}
\label{fig:sims}
\end{figure}

Results are shown in Figure \ref{fig:sims}.
Red dashed lines mark the true failure probabilities, and grey shaded regions highlight 
$\{\alpha\pm2\sigma_\alpha\}$ where $\sigma_\alpha$ is calculated from the true $\alpha$ and 
chosen $M$.  This region provides important context -- it captures the intrinsic
variability that would accompany any MC estimator.  In other words,
even if we used the true simulator for all $\{\mathbf{x}_i\}_{i=1}^M$ samples (\ref{eq:mc}), 
we would obtain a boxplot roughly spanning the grey interval.  We thus 
consider any results within this region to be effective estimates.  We 
could tighten the interval (and the boxplots of the MC methods) 
by increasing $M$ without requiring any additional 
simulator evaluations, another advantage MC holds over IS.

Across the board, both importance sampling based methods (``SIIS'' and ``SIIS UCB'')
suffered from high variability in their failure probability estimates.  Although
their median performance was generally on target (indicative of their unbiasedness),
these methods consistently offered the poorest performance.  The more conservative 
``SIIS UCB'' had no apparent advantage over the traditional ``SIIS.'' We believe importance
sampling is simply ill-suited for our low data settings.  In all scenarios aside
from the simplest two-dimensional Herbie function (where performance was comparable), 
our two-stage design (``Two-stage Entropy'') outperformed the exhaustive contour 
location approach (``Exhaustive CL'').  Particularly
as dimension grows and the response surface becomes more complex, there is a clear
advantage to our two-stage design which transitions from CL to exploitation of MC
samples.  Within the two-stage designs, there was generally agreement between the 
two variations we considered  for allocating samples to $\mathcal{U}$ (``Entropy''
and ``Proximity''), but when there was disagreement the entropy criterion offered
higher accuracy, with uniformly superior worst-case performance.  As suspected, the entropy
criterion which incorporates uncertainty is more effective at identifying samples which the 
surrogate would misclassify.

\section{RAE-2822 airfoil computer experiment}\label{sec:airfoil}

The RAE-2822 airfoil computer experiment simulates the flow of air around
an aircraft wing.  The simulation suite is solved via SU2, and the software is
publicly available \citep{economon2016su2}.  We consider seven inputs:
four shape parameters ($S_i$ for $i=1,\dots,4$) and 3 environmental parameters, namely
angle of attack, Reynolds number, and Mach number.  We are interested in the lift 
over drag ($L/D$) ratio.  $L/D$ values that are too low represent inefficient 
aircraft performance.  We define a failure threshold at $t=3$ such that $L/D < 3$ 
indicates failure.  [Note the change from $f(\mathbf{x})>t$ to $f(\mathbf{x})<t$ does
not affect the implementation of our method.]  This computer experiment requires
nearly 30 minutes of compute time with serial execution, but reductions are possible
with parallel execution.

Our objective is to estimate the probability of failure
given the following input distribution:
\[
\begin{aligned}
S_i &\sim \mathrm{Unif}(-1\times 10^{-4}, 1\times 10^{-4}) \;\;\textrm{for}\;\; i = 1, \dots, 4 \\
\textrm{Angle of Attack} &\sim \mathcal{N}\left(\mu = 5, \sigma = 1\right)
\;\;\textrm{truncated to}\;\;[0,10]\\
\textrm{Reynolds number} &\sim \mathcal{N}\left(\mu = 1\times 10^7, \sigma = 1\times 10^6\right)
\;\;\textrm{truncated to}\;\;[5\times 10^6,1.5\times 10^7]\\
\textrm{Mach number} &\sim \mathcal{N}\left(\mu = 0.8, \sigma = 0.02\right)
\;\;\textrm{truncated to}\;\;[0.7,0.9].\\
\end{aligned}
\]
We fix our total budget at $N=500$ simulator evaluations and use $M=2.5\times 10^6$ for the
size of $X_M$.
\citet{booth2024contour} found that DGP surrogates outperformed 
GP surrogates for contour location of this experiment.  We follow the settings
therein, starting with an LHS sample of size $n_0 = 100$ and fitting a Bayesian DGP
surrogate with the {\tt deepgp} package.  We proceed with contour location (Stage 1)
following \citet{booth2024contour}, stopping after every 10th acquisition to estimate
$\hat{\alpha}$ and check whether it is within the $\pm\hat{\sigma}_\alpha$ bounds.
Once we halt contour location (Section \ref{sec:stop}), we proceed to Stage 2, using 
entropy to inform our hybrid MC estimator (Section \ref{sec:hybmc}).
Given the larger input dimension, we require at least 40 observed failures before we 
consider halting surrogate training.  Since the true failure probability is unknown,
we opt to repeat this entire exercise twice, with a newly randomized starting design.  
We consider the agreement between the two randomized repetitions
as an indication that our procedure is well-calibrated. 

\begin{figure}[h!]
\centering
\includegraphics[width=0.8\linewidth]{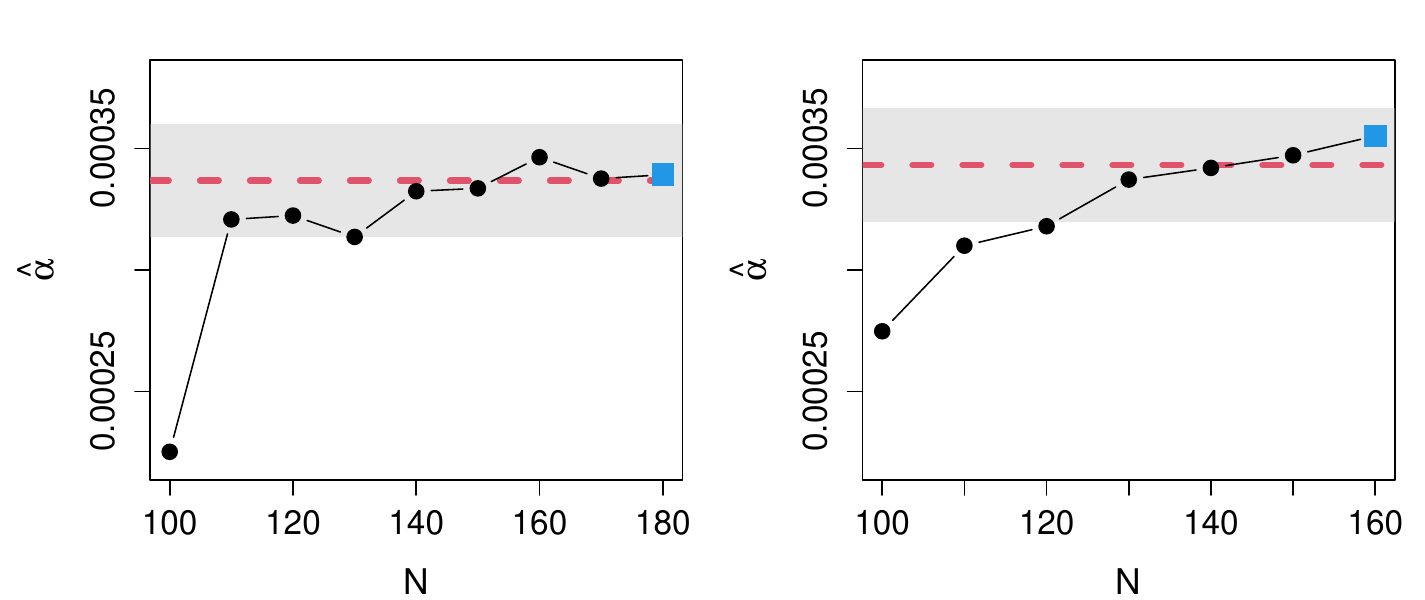}
\caption{Progress in $\hat{\alpha}$ across two contour locating designs for the airfoil
computer experiment.  Proposed two-stage failure probability estimate shown in dashed red.
Grey shading indicates $\hat{\alpha}\pm 2\hat{\sigma}_\alpha$ for these final estimates. 
Left/right panels show two repetitions with newly randomized starting designs.}
\label{fig:airfoil}
\end{figure}

Figure \ref{fig:airfoil} reports the progress in $\hat{\alpha}$ over the course of 
each design.  Our stopping criterion halted the sequential designs at $n=180$ and $n=160$,
leaving $B=320$ and $B=340$ for the second design stage, respectively.  
Our two hybrid Monte Carlo estimates were 
$\hat{\alpha}_\textrm{HYB} = \{0.0003368, 0.0003432\}$.  These values 
are indicated by the red dashed lines in Figure \ref{fig:airfoil}.  The grey shading
shows $\hat{\alpha}\pm 2\hat{\sigma}_\alpha$, calculated from these final estimates.  
Although the two estimates vary slightly, they are both within the bounds of reasonable 
variability for our chosen $M$.  Additionally, notice how the final hybrid MC estimate
of the second exercise (right panel) brought the failure probability lower than
what the surrogate had originally estimated, closer to the estimate of our first attempt
(left panel).  We take this as an indication that the hybrid estimator of Stage 2 is effectively
adjusting for surrogate inaccuracies remaining after Stage 1.

\section{Discussion}\label{sec:discuss}

We presented a unique two-stage surrogate design for estimating failure probabilities
of expensive computer experiments, which outperforms both
exhaustive contour location and surrogate-informed importance sampling. 
Starting with a contour locating sequential design,
we halt contour location once surrogate training has plateaued, as indicated by practical convergence
of a surrogate Monte Carlo estimator.  Then we use the remaining budget of simulator evaluations
on the Monte Carlo samples with highest entropy, to ultimately return a hybrid MC estimate
of $\alpha$.  Our design strategically leverages the uncertainty 
quantification provided by the surrogate to reserve expensive simulator evaluations
for samples with high classification uncertainty.  While we limited our experiments
to GP and DGP surrogates, our proposed scheme
is applicable to any surrogate which provides posterior predictive distributions 
(which are required for entropy calculations).  Extensions to other surrogates, 
such a treed Gaussian processes \citep{gramacy2008bayesian} or Bayesian
additive regression trees \citep{chipman2010bart}, is straightforward.

Our two-stage design is computationally efficient in the number of
evaluations of the expensive simulator, but it requires many predictive 
evaluations from the surrogate.  While surrogate predictions are very cheap, 
collecting billions of them is not necessarily trivial.
The computational cost of obtaining these predictions increases
with $n/N$, $d$, and $M$, but it also depends on the surrogate and computing environment.
For context, the most expensive predictions we needed in our exercises were those 
of our motivating airfoil simulator, with the 
fully-Bayesian DGP trained on all $N=500$ points in 7-dimensions.
With parallelization over 10 CPUs, it took just over 10 minutes to obtain one million 
predictions.  For smaller $\alpha$ requiring $M$ in the billions, the cost of
these surrogate predictions may present an obstacle, but it is still far cheaper than
comparative evaluations of the true expensive computer simulator.

There may be avenues to circumvent surrogate predictions at all $X_M$ locations.
We hypothesize that a classification rule based on nearest neighbors could
work well: for a particular $\mathbf{x}_i$, if the surrogate predicts that all of its $k$-many 
nearest neighbors are ``passes,'' then we could presume $\mathbbm{1}_{\{\mu_n(\mathbf{x}_i) > t\}}=0$
without needing to obtain $\mu_n(\mathbf{x}_i)$ directly.  This requires effective tuning
of $k$, and would only speed-up computations if the cost of obtaining the nearest neighbor
sets is less than that of obtaining the single surrogate prediction.  For these reasons,
we did not entertain such workarounds in our current work.

In all of our examples, our stopping criterion terminated contour location
before the complete budget ($N$) was exhausted.  If a simulation budget is severely 
restricted, the budget could be exhausted before our proposed stopping criterion is met.  
While we prefer to have additional evaluations reserved for Stage 2, 
we contend that using the entire budget for contour location in Stage 1 to inform a 
traditional surrogate MC estimate is the best option in this scenario.  As we've shown, 
importance sampling based methods perform poorly when data is limited.

We have exclusively focused on deterministic computer experiments.  In fact, the noise-free
nature of the simulator is what enabled us to frame our contribution as a two-stage
design.  Once $f(\mathbf{x}_i)$ has been observed and incorporated into the surrogate, 
surrogate predictions should interpolate these observations.  Thus our hybrid MC 
estimator (\ref{eq:hybmc}) is really a surrogate MC estimator (\ref{eq:alphasurr}),
with a strategically trained surrogate.  
Stochastic simulators \citep[e.g.,][]{baker2022analyzing} present a unique
challenge as repeated observations of noisy $f(\mathbf{x})$ may result in a mix of
passes and failures for the same $\mathbf{x}$.  To accommodate this, we could
have broadened the formulation of Eq.~(\ref{eq:problem}) by replacing the indicator
function $\mathbbm{1}_{\{f(\mathbf{x}) > t\}}$ with the unknown probability
$\mathrm{P}(f(\mathbf{x}) > t)$, which recovers the indicator when $f$ is deterministic.  
The key advantage of our hybrid estimator -- that observing $f(\mathbf{x}_i)$ ensures
correct classification of $\mathbbm{1}_{\{f(\mathbf{x}_i)>t\}}$ -- is not applicable
to noisy simulators.  Instead, new active learning strategies that strategically 
allocate replicates \citep{binois2019replication} and inform a MC estimator 
leveraging surrogate posterior predictive probabilities are warranted.  
We leave this as an avenue for future work.

When the input distribution is not uniform over $\mathcal{X}$, it could be 
beneficial to leverage information about $p(\mathbf{x})$ in the sequential
training of the surrogate \citep{abdelmalek2024bayesian}.  There is arguably no need to 
train a surrogate in a region where
inputs will never occur.  Incorporating this information
in the contour location stage could improve failure probability estimation 
downstream, but we have seen positive impacts from having ``anchor points'' in regions
of very low density (such as those around the edges in 
Figure \ref{fig:herbie}).  We suspect there is a delicate balance that 
would work well here, but this avenue is currently underexplored.
Furthermore, while we focused on stochastic inputs, it is possible to 
extend our approach to incorporate controllable/design variables by
using point masses in $p(\mathbf{x})$.

Finally, estimation of failure probabilities is often an intermediate step within a 
larger objective, such as a stochastically constrained optimization.  For 
example, in the context of our airfoil experiment, we may wish to find the optimal
wing configuration given that the probability of failure remains below
a specified threshold.  The ability to effectively estimate failure 
probabilities from limited data, which we have tackled here, is integral to 
the success of this larger enterprise.  While we focused on allocating
a predetermined budget of evaluations, it could be useful to develop a decision
rule for halting Stage 2, potentially reserving some evaluations, particularly 
if this design is within a larger optimization loop.

\singlespacing

\bibliographystyle{jasa}
\bibliography{ref}

\begin{thebibliography}{69}
\newcommand{\enquote}[1]{``#1''}
\expandafter\ifx\csname natexlab\endcsname\relax\def\natexlab#1{#1}\fi

\bibitem[\protect\citename{Abdelmalek-Lomenech et~al.,
  }2024]{abdelmalek2024bayesian}
Abdelmalek-Lomenech, R.~A., Bect, J., Chabridon, V., and Vazquez, E. (2024).
\newblock \enquote{Bayesian sequential design of computer experiments for
  quantile set inversion.}
\newblock {\em Technometrics\/}, , just-accepted, 1--14.

\bibitem[\protect\citename{Au and Beck, }2001]{au2001estimation}
Au, S.-K. and Beck, J.~L. (2001).
\newblock \enquote{Estimation of small failure probabilities in high dimensions
  by subset simulation.}
\newblock {\em Probabilistic engineering mechanics\/}, 16, 4, 263--277.

\bibitem[\protect\citename{Azzimonti et~al., }2021]{azzimonti2021adaptive}
Azzimonti, D., Ginsbourger, D., Chevalier, C., Bect, J., and Richet, Y. (2021).
\newblock \enquote{Adaptive design of experiments for conservative estimation
  of excursion sets.}
\newblock {\em Technometrics\/}, 63, 1, 13--26.

\bibitem[\protect\citename{Baker et~al., }2022]{baker2022analyzing}
Baker, E., Barbillon, P., Fadikar, A., Gramacy, R.~B., Herbei, R., Higdon, D.,
  Huang, J., Johnson, L.~R., Ma, P., Mondal, A., et~al. (2022).
\newblock \enquote{Analyzing stochastic computer models: A review with
  opportunities.}
\newblock {\em Statistical Science\/}, 37, 1, 64--89.

\bibitem[\protect\citename{Batet et~al., }2014]{batet2014modelling}
Batet, L., Alvarez-Fernandez, J.~M., de~les Valls, E.~M., Martinez-Quiroga, V.,
  Perez, M., Reventos, F., and Sedano, L. (2014).
\newblock \enquote{Modelling of a supercritical CO2 power cycle for nuclear
  fusion reactors using RELAP5--3D.}
\newblock {\em Fusion Engineering and Design\/}, 89, 4, 354--359.

\bibitem[\protect\citename{Bect et~al., }2012]{bect2012sequential}
Bect, J., Ginsbourger, D., Li, L., Picheny, V., and Vazquez, E. (2012).
\newblock \enquote{Sequential design of computer experiments for the estimation
  of a probability of failure.}
\newblock {\em Statistics and Computing\/}, 22, 773--793.

\bibitem[\protect\citename{Belot et~al., }2021]{belot2021impact}
Belot, I., Vidal, D., Greiner, R., Votsmeier, M., Hayes, R.~E., and Bertrand,
  F. (2021).
\newblock \enquote{Impact of washcoat distribution on the catalytic performance
  of gasoline particulate filters as predicted by lattice Boltzmann
  simulations.}
\newblock {\em Chemical Engineering Journal\/}, 406, 127040.

\bibitem[\protect\citename{Bichon et~al., }2008]{bichon2008efficient}
Bichon, B.~J., Eldred, M.~S., Swiler, L.~P., Mahadevan, S., and McFarland,
  J.~M. (2008).
\newblock \enquote{Efficient global reliability analysis for nonlinear implicit
  performance functions.}
\newblock {\em AIAA journal\/}, 46, 10, 2459--2468.

\bibitem[\protect\citename{Binois et~al., }2019]{binois2019replication}
Binois, M., Huang, J., Gramacy, R.~B., and Ludkovski, M. (2019).
\newblock \enquote{Replication or exploration? Sequential design for stochastic
  simulation experiments.}
\newblock {\em Technometrics\/}, 61, 1, 7--23.

\bibitem[\protect\citename{Booth, }2024]{deepgp}
Booth, A.~S. (2024).
\newblock {\em deepgp: Bayesian Deep Gaussian Processes using MCMC\/}.
\newblock R package version 1.1.3.

\bibitem[\protect\citename{Booth et~al.,
  }2024{\natexlab{a}}]{booth2024nonstationary}
Booth, A.~S., Cooper, A., and Gramacy, R.~B. (2024{\natexlab{a}}).
\newblock \enquote{Nonstationary Gaussian process surrogates.}
\newblock {\em arXiv:2305.19242\/}.

\bibitem[\protect\citename{Booth et~al.,
  }2024{\natexlab{b}}]{booth2024actively}
Booth, A.~S., Gramacy, R., and Renganathan, A. (2024{\natexlab{b}}).
\newblock \enquote{Actively learning deep Gaussian process models for failure
  contour and probability estimation.}
\newblock In {\em AIAA SCITECH 2024 Forum\/},  0577.

\bibitem[\protect\citename{Booth et~al., }2024{\natexlab{c}}]{booth2024contour}
Booth, A.~S., Renganathan, S.~A., and Gramacy, R.~B. (2024{\natexlab{c}}).
\newblock \enquote{Contour Location for Reliability in Airfoil Simulation
  Experiments using Deep Gaussian Processes.}
\newblock {\em Annals of Applied Statistics\/}, , just-accepted.
\newblock ArXiv:2308.04420.

\bibitem[\protect\citename{Bugallo et~al., }2017]{bugallo2017adaptive}
Bugallo, M.~F., Elvira, V., Martino, L., Luengo, D., Miguez, J., and Djuric,
  P.~M. (2017).
\newblock \enquote{Adaptive importance sampling: The past, the present, and the
  future.}
\newblock {\em IEEE Signal Processing Magazine\/}, 34, 4, 60--79.

\bibitem[\protect\citename{Cheng and Lu, }2020]{cheng2020structural}
Cheng, K. and Lu, Z. (2020).
\newblock \enquote{Structural reliability analysis based on ensemble learning
  of surrogate models.}
\newblock {\em Structural Safety\/}, 83, 101905.

\bibitem[\protect\citename{Cheng et~al., }2023]{cheng2023rare}
Cheng, K., Papaioannou, I., Lu, Z., Zhang, X., and Wang, Y. (2023).
\newblock \enquote{Rare event estimation with sequential directional importance
  sampling.}
\newblock {\em Structural Safety\/}, 100, 102291.

\bibitem[\protect\citename{Chevalier et~al., }2014]{chevalier2014fast}
Chevalier, C., Bect, J., Ginsbourger, D., Vazquez, E., Picheny, V., and Richet,
  Y. (2014).
\newblock \enquote{Fast parallel kriging-based stepwise uncertainty reduction
  with application to the identification of an excursion set.}
\newblock {\em Technometrics\/}, 56, 4, 455--465.

\bibitem[\protect\citename{Chipman et~al., }2010]{chipman2010bart}
Chipman, H.~A., George, E.~I., and McCulloch, R.~E. (2010).
\newblock \enquote{{BART: Bayesian additive regression trees}.}
\newblock {\em The Annals of Applied Statistics\/}, 4, 1, 266 -- 298.

\bibitem[\protect\citename{Cole et~al., }2023]{cole2023entropy}
Cole, D.~A., Gramacy, R.~B., Warner, J.~E., Bomarito, G.~F., Leser, P.~E., and
  Leser, W.~P. (2023).
\newblock \enquote{Entropy-based adaptive design for contour finding and
  estimating reliability.}
\newblock {\em Journal of Quality Technology\/}, 55, 1, 43--60.

\bibitem[\protect\citename{Dalbey and Swiler, }2014]{dalbey2014gaussian}
Dalbey, K.~R. and Swiler, L.~P. (2014).
\newblock \enquote{Gaussian process adaptive importance sampling.}
\newblock {\em International Journal for Uncertainty Quantification\/}, 4, 2.

\bibitem[\protect\citename{Damianou and Lawrence, }2013]{damianou2013deep}
Damianou, A. and Lawrence, N.~D. (2013).
\newblock \enquote{Deep gaussian processes.}
\newblock In {\em Artificial intelligence and statistics\/},  207--215. PMLR.

\bibitem[\protect\citename{Dubourg et~al., }2011]{dubourg2011reliability}
Dubourg, V., Sudret, B., and Bourinet, J.-M. (2011).
\newblock \enquote{Reliability-based design optimization using kriging
  surrogates and subset simulation.}
\newblock {\em Structural and Multidisciplinary Optimization\/}, 44, 673--690.

\bibitem[\protect\citename{Duhamel et~al., }2023]{duhamel2023version}
Duhamel, C., Helbert, C., Munoz~Zuniga, M., Prieur, C., and Sinoquet, D.
  (2023).
\newblock \enquote{A SUR version of the Bichon criterion for excursion set
  estimation.}
\newblock {\em Statistics and Computing\/}, 33, 2, 41.

\bibitem[\protect\citename{Echard et~al., }2011]{echard2011ak}
Echard, B., Gayton, N., and Lemaire, M. (2011).
\newblock \enquote{AK-MCS: an active learning reliability method combining
  Kriging and Monte Carlo simulation.}
\newblock {\em Structural Safety\/}, 33, 2, 145--154.

\bibitem[\protect\citename{Economon et~al., }2016]{economon2016su2}
Economon, T.~D., Palacios, F., Copeland, S.~R., Lukaczyk, T.~W., and Alonso,
  J.~J. (2016).
\newblock \enquote{SU2: An open-source suite for multiphysics simulation and
  design.}
\newblock {\em Aiaa Journal\/}, 54, 3, 828--846.

\bibitem[\protect\citename{Fuhrl{\"a}nder and Sch{\"o}ps,
  }2020]{fuhrlander2020blackbox}
Fuhrl{\"a}nder, M. and Sch{\"o}ps, S. (2020).
\newblock \enquote{A blackbox yield estimation workflow with Gaussian process
  regression applied to the design of electromagnetic devices.}
\newblock {\em Journal of Mathematics in Industry\/}, 10, 1, 25.

\bibitem[\protect\citename{Gramacy, }2020]{gramacy2020surrogates}
Gramacy, R.~B. (2020).
\newblock {\em Surrogates: Gaussian Process Modeling, Design, and Optimization
  for the Applied Sciences\/}.
\newblock CRC press.

\bibitem[\protect\citename{Gramacy and Apley, }2015]{gramacy2015local}
Gramacy, R.~B. and Apley, D.~W. (2015).
\newblock \enquote{Local Gaussian process approximation for large computer
  experiments.}
\newblock {\em Journal of Computational and Graphical Statistics\/}, 24, 2,
  561--578.

\bibitem[\protect\citename{Gramacy and Lee, }2008]{gramacy2008bayesian}
Gramacy, R.~B. and Lee, H. K.~H. (2008).
\newblock \enquote{Bayesian treed Gaussian process models with an application
  to computer modeling.}
\newblock {\em Journal of the American Statistical Association\/}, 103, 483,
  1119--1130.

\bibitem[\protect\citename{Gramacy et~al., }2022]{gramacy2022triangulation}
Gramacy, R.~B., Sauer, A., and Wycoff, N. (2022).
\newblock \enquote{Triangulation candidates for bayesian optimization.}
\newblock {\em Advances in Neural Information Processing Systems\/}, 35,
  35933--35945.

\bibitem[\protect\citename{Haldar and Mahadevan, }1995]{haldar1995first}
Haldar, A. and Mahadevan, S. (1995).
\newblock \enquote{First-order and second-order reliability methods.}
\newblock In {\em Probabilistic Structural Mechanics Handbook: theory and
  industrial applications\/},  27--52. Springer.

\bibitem[\protect\citename{Hristov et~al., }2019]{hristov2019adaptive}
Hristov, P., DiazDelaO, F., Farooq, U., and Kubiak, K. (2019).
\newblock \enquote{Adaptive Gaussian process emulators for efficient
  reliability analysis.}
\newblock {\em Applied Mathematical Modelling\/}, 71, 138--151.

\bibitem[\protect\citename{Huang et~al., }2016]{huang2016assessing}
Huang, X., Chen, J., and Zhu, H. (2016).
\newblock \enquote{Assessing small failure probabilities by AK--SS: An active
  learning method combining Kriging and Subset Simulation.}
\newblock {\em Structural Safety\/}, 59, 86--95.

\bibitem[\protect\citename{Ishigami and Homma, }1990]{ishigami1990importance}
Ishigami, T. and Homma, T. (1990).
\newblock \enquote{An importance quantification technique in uncertainty
  analysis for computer models.}
\newblock In {\em [1990] Proceedings. First international symposium on
  uncertainty modeling and analysis\/},  398--403. IEEE.

\bibitem[\protect\citename{Joseph, }2016]{joseph2016space}
Joseph, V.~R. (2016).
\newblock \enquote{Space-filling designs for computer experiments: A review.}
\newblock {\em Quality Engineering\/}, 28, 1, 28--35.

\bibitem[\protect\citename{Kurtz and Song, }2013]{kurtz2013cross}
Kurtz, N. and Song, J. (2013).
\newblock \enquote{Cross-entropy-based adaptive importance sampling using
  Gaussian mixture.}
\newblock {\em Structural Safety\/}, 42, 35--44.

\bibitem[\protect\citename{Lee et~al., }2011]{lee2011optimization}
Lee, H., Gramacy, R., Linkletter, C., and Gray, G. (2011).
\newblock \enquote{Optimization subject to hidden constraints via statistical
  emulation.}
\newblock {\em Pacific Journal of Optimization\/}, 7, 3, 467--478.

\bibitem[\protect\citename{Li and Xiu, }2010]{li2010evaluation}
Li, J. and Xiu, D. (2010).
\newblock \enquote{Evaluation of failure probability via surrogate models.}
\newblock {\em Journal of Computational Physics\/}, 229, 23, 8966--8980.

\bibitem[\protect\citename{Li et~al., }2021]{li2021ass}
Li, M., Wang, G., Qian, L., Li, X., and Ma, Z. (2021).
\newblock \enquote{ASS-GPR: Adaptive Sequential Sampling Method Based on
  Gaussian Process Regression for Reliability Analysis of Complex Geotechnical
  Engineering.}
\newblock {\em International Journal of Geomechanics\/}, 21, 10, 04021192.

\bibitem[\protect\citename{Lu et~al., }2023]{lu2023agp}
Lu, N., Li, Y.-F., Huang, H.-Z., Mi, J., and Niazi, S.~G. (2023).
\newblock \enquote{AGP-MCS+ D: An active learning reliability analysis method
  combining dependent Gaussian process and Monte Carlo simulation.}
\newblock {\em Reliability Engineering \& System Safety\/}, 240, 109541.

\bibitem[\protect\citename{MacKay, }1995]{mackay1995bayesian}
MacKay, D.~J. (1995).
\newblock \enquote{Bayesian neural networks and density networks.}
\newblock {\em Nuclear Instruments and Methods in Physics Research Section A:
  Accelerators, Spectrometers, Detectors and Associated Equipment\/}, 354, 1,
  73--80.

\bibitem[\protect\citename{Marques et~al., }2018]{marques2018contour}
Marques, A., Lam, R., and Willcox, K. (2018).
\newblock \enquote{Contour location via entropy reduction leveraging multiple
  information sources.}
\newblock {\em Advances in neural information processing systems\/}, 31.

\bibitem[\protect\citename{McKay et~al., }2000]{mckay2000comparison}
McKay, M.~D., Beckman, R.~J., and Conover, W.~J. (2000).
\newblock \enquote{A comparison of three methods for selecting values of input
  variables in the analysis of output from a computer code.}
\newblock {\em Technometrics\/}, 42, 1, 55--61.

\bibitem[\protect\citename{Mia et~al., }2017]{mia2017modal}
Mia, M.~S., Islam, M.~S., and Ghosh, U. (2017).
\newblock \enquote{Modal analysis of cracked cantilever beam by finite element
  simulation.}
\newblock {\em Procedia engineering\/}, 194, 509--516.

\bibitem[\protect\citename{Ming et~al., }2023]{ming2023deep}
Ming, D., Williamson, D., and Guillas, S. (2023).
\newblock \enquote{Deep Gaussian process emulation using stochastic
  imputation.}
\newblock {\em Technometrics\/}, 65, 2, 150--161.

\bibitem[\protect\citename{Oh and Berger, }1992]{oh1992adaptive}
Oh, M.-S. and Berger, J.~O. (1992).
\newblock \enquote{Adaptive importance sampling in Monte Carlo integration.}
\newblock {\em Journal of statistical computation and simulation\/}, 41, 3-4,
  143--168.

\bibitem[\protect\citename{O'Hagan, }1987]{ohagan1987monte}
O'Hagan, A. (1987).
\newblock \enquote{Monte Carlo is fundamentally unsound.}
\newblock {\em The Statistician\/},  247--249.

\bibitem[\protect\citename{Park et~al., }2017]{park2017remarks}
Park, C., Haftka, R.~T., and Kim, N.~H. (2017).
\newblock \enquote{Remarks on multi-fidelity surrogates.}
\newblock {\em Structural and Multidisciplinary Optimization\/}, 55,
  1029--1050.

\bibitem[\protect\citename{Pedregosa et~al., }2011]{pedregosa2011scikit}
Pedregosa, F., Varoquaux, G., Gramfort, A., Michel, V., Thirion, B., Grisel,
  O., Blondel, M., Prettenhofer, P., Weiss, R., Dubourg, V., et~al. (2011).
\newblock \enquote{Scikit-learn: Machine learning in Python.}
\newblock {\em the Journal of machine Learning research\/}, 12, 2825--2830.

\bibitem[\protect\citename{Peherstorfer et~al.,
  }2016]{peherstorfer2016multifidelity}
Peherstorfer, B., Cui, T., Marzouk, Y., and Willcox, K. (2016).
\newblock \enquote{Multifidelity importance sampling.}
\newblock {\em Computer Methods in Applied Mechanics and Engineering\/}, 300,
  490--509.

\bibitem[\protect\citename{Picheny et~al., }2013]{picheny2013benchmark}
Picheny, V., Wagner, T., and Ginsbourger, D. (2013).
\newblock \enquote{A benchmark of kriging-based infill criteria for noisy
  optimization.}
\newblock {\em Structural and multidisciplinary optimization\/}, 48, 607--626.

\bibitem[\protect\citename{Rajaram et~al., }2021]{rajaram2021empirical}
Rajaram, D., Puranik, T.~G., Ashwin~Renganathan, S., Sung, W., Fischer, O.~P.,
  Mavris, D.~N., and Ramamurthy, A. (2021).
\newblock \enquote{Empirical assessment of deep gaussian process surrogate
  models for engineering problems.}
\newblock {\em Journal of Aircraft\/}, 58, 1, 182--196.

\bibitem[\protect\citename{Ranjan et~al., }2008]{ranjan2008sequential}
Ranjan, P., Bingham, D., and Michailidis, G. (2008).
\newblock \enquote{Sequential experiment design for contour estimation from
  complex computer codes.}
\newblock {\em Technometrics\/}, 50, 4, 527--541.

\bibitem[\protect\citename{Rasmussen et~al., }2006]{rasmussen2006gaussian}
Rasmussen, C.~E., Williams, C.~K., et~al. (2006).
\newblock {\em Gaussian Processes for Machine Learning\/}, vol.~1.
\newblock Springer.

\bibitem[\protect\citename{Renganathan, }2024]{renganathan2024efficient}
Renganathan, A. (2024).
\newblock \enquote{Efficient reliability analysis with multifidelity Gaussian
  processes and normalizing flows.}
\newblock In {\em AIAA SCITECH 2024 Forum\/},  0576.

\bibitem[\protect\citename{Renganathan et~al., }2023]{renganathan2023camera}
Renganathan, S.~A., Rao, V., and Navon, I.~M. (2023).
\newblock \enquote{CAMERA: A method for cost-aware, adaptive, multifidelity,
  efficient reliability analysis.}
\newblock {\em Journal of Computational Physics\/}, 472, 111698.

\bibitem[\protect\citename{Reynolds et~al., }2009]{reynolds2009gaussian}
Reynolds, D.~A. et~al. (2009).
\newblock \enquote{Gaussian mixture models.}
\newblock {\em Encyclopedia of biometrics\/}, 741, 659-663.

\bibitem[\protect\citename{Santner et~al., }2018]{santner2003design}
Santner, T.~J., Williams, B.~J., and Notz, W.~I. (2018).
\newblock {\em The Design and Analysis of Computer Experiments\/}.
\newblock Springer.

\bibitem[\protect\citename{Sauer, }2023]{sauer2023deep}
Sauer, A. (2023).
\newblock \enquote{Deep Gaussian process surrogates for computer experiments.}
\newblock Ph.D. thesis, Virginia Polytechnic Institute and State University.

\bibitem[\protect\citename{Sauer et~al., }2023{\natexlab{a}}]{sauer2023vecchia}
Sauer, A., Cooper, A., and Gramacy, R.~B. (2023{\natexlab{a}}).
\newblock \enquote{Vecchia-approximated deep Gaussian processes for computer
  experiments.}
\newblock {\em Journal of Computational and Graphical Statistics\/}, 32, 3,
  824--837.

\bibitem[\protect\citename{Sauer et~al., }2023{\natexlab{b}}]{sauer2023active}
Sauer, A., Gramacy, R.~B., and Higdon, D. (2023{\natexlab{b}}).
\newblock \enquote{Active learning for deep Gaussian process surrogates.}
\newblock {\em Technometrics\/}, 65, 4--18.

\bibitem[\protect\citename{Scrucca et~al., }2016]{mclust}
Scrucca, L., Fop, M., Murphy, T.~B., and Raftery, A.~E. (2016).
\newblock \enquote{{mclust} 5: clustering, classification and density
  estimation using {G}aussian finite mixture models.}
\newblock {\em The {R} Journal\/}, 8, 1, 289--317.

\bibitem[\protect\citename{Srinivasan, }2002]{srinivasan2002importance}
Srinivasan, R. (2002).
\newblock {\em Importance Sampling: Applications in Communications and
  Detection\/}.
\newblock Springer Science \& Business Media.

\bibitem[\protect\citename{Su et~al., }2017]{su2017gaussian}
Su, G., Peng, L., and Hu, L. (2017).
\newblock \enquote{A Gaussian process-based dynamic surrogate model for complex
  engineering structural reliability analysis.}
\newblock {\em Structural Safety\/}, 68, 97--109.

\bibitem[\protect\citename{Tabandeh et~al., }2022]{tabandeh2022review}
Tabandeh, A., Jia, G., and Gardoni, P. (2022).
\newblock \enquote{A review and assessment of importance sampling methods for
  reliability analysis.}
\newblock {\em Structural Safety\/}, 97, 102216.

\bibitem[\protect\citename{Tokdar and Kass, }2010]{tokdar2010importance}
Tokdar, S.~T. and Kass, R.~E. (2010).
\newblock \enquote{Importance sampling: a review.}
\newblock {\em Wiley Interdisciplinary Reviews: Computational Statistics\/}, 2,
  1, 54--60.

\bibitem[\protect\citename{Yazdi, }2022]{yazdi2022fast}
Yazdi, F. (2022).
\newblock \enquote{Fast deep gaussian process modeling and design for large
  complex computer experiments.}
\newblock Ph.D. thesis, Simon Fraser University.

\bibitem[\protect\citename{Zhang et~al., }2015]{zhang2015efficient}
Zhang, L., Lu, Z., and Wang, P. (2015).
\newblock \enquote{Efficient structural reliability analysis method based on
  advanced Kriging model.}
\newblock {\em Applied Mathematical Modelling\/}, 39, 2, 781--793.

\bibitem[\protect\citename{Zhu and Du, }2016]{zhu2016reliability}
Zhu, Z. and Du, X. (2016).
\newblock \enquote{Reliability analysis with Monte Carlo simulation and
  dependent Kriging predictions.}
\newblock {\em Journal of Mechanical Design\/}, 138, 12, 121403.

\end{thebibliography}

\newpage
\begin{center}
{\LARGE\bf SUPPLEMENTARY MATERIAL}
\end{center}
\appendix

\section{Test functions}\label{supp:functions}

Here we provide the details of the functions, thresholds, and input
distributions used in the benchmark exercises of Section \ref{sec:results}.
In each case, failures are defined as $f(\mathbf{x}) > t$.

\paragraph{Herbie.}The 2d Herbie function \citep{lee2011optimization} is 
defined in over $[-2, 2]^2$ as 
\[
f(\mathbf{x}) = \prod_{i=1}^2 \mathrm{exp}\left(-(x_i-1)^2\right) + 
    \mathrm{exp}\left(-0.8*(x_i+1)^2\right) - 0.05*\sin(8*(x_i+1)).
\]
We set the failure threshold at $t=1.065$ and define the input distribution 
as $x_i \stackrel{\mathrm{ind}}{\sim} \mathcal{N}(0, 0.36)$ for $i=1,2$,
truncated to $[-2, 2]$.

\paragraph{Ishigami.}The 3d Ishigami function \citep{ishigami1990importance} 
is defined over $[-\pi, \pi]^3$ as 
\[
f(\mathbf{x}) =\sin(x_1) + 5*\sin(x_2)^2 + 0.1*x_3^4*\sin(x_1).
\]
We set the failure threshold at $t=10.244$ and define the input distribution 
as 
\[
\begin{aligned}
x_1&\sim\mathcal{N}(-1, 1) \;\;\textrm{truncated to}\;\;[-\pi,\pi] \\
x_2&\sim\mathcal{N}(1.5, 1.5) \;\;\textrm{truncated to}\;\;[-\pi,\pi] \\
x_3&\sim\mathrm{Uniform}(-\pi, \pi).
\end{aligned}
\]

\paragraph{Hartmann.}The 6d Hartmann function \citep{picheny2013benchmark} 
is defined over $[0, 1]^6$ as 
\[
f(\mathbf{x}) = -\sum_{i=1}^4\alpha_i\mathrm{exp}\left(-\sum_{j=1}^6 
    A_{ij}(x_j - P_{ij})^2\right)\;\;\textrm{where}
\]
\[
\alpha^\top= \begin{bmatrix} 1 \\ 1.2 \\ 3 \\ 3.2 \end{bmatrix}
\quad\quad
A^\top = \begin{bmatrix} 10 & .05 & 3 & 17 \\
                  3 & 10 & 3.5 & 8 \\
                  17 & 17 & 1.7 & .05 \\ 
                  3.5 & .1 & 10 & 10 \\ 
                  7 & 8 & 17 & .1 \\ 
                  8 & 14 & 8 & 14 \end{bmatrix}
\quad\textrm{and}\quad
P^\top = \begin{bmatrix} .1312 & .2329 & .2348 & .4047 \\ 
                  .1696 & .4135 & .1451 & .8828 \\
                  .5569 & .8307 & .3522 & .8732 \\ 
                  .0124 & .3736 & .2883 & .5743 \\
                  .8283 & .1004 & .3047 & .1091 \\ 
                  .5886 & .9991 & .6650 & .0381\end{bmatrix}.
\]
We set the failure threshold at $t=2.46$ and define the input distribution 
as $x_i \stackrel{\mathrm{ind}}{\sim} \mathcal{N}(0.5, 0.1)$ for $i=1,\dots,6$,
truncated to $[0, 1]$.

\paragraph{Plateau.}The Plateau function \citep{booth2024contour} is defined
in arbitrary dimension over $[0,1]^d$ as
\[
f(\mathbf{x}) = 2*\Phi\left[\sqrt{2}\left(-4-3\sum_{i=1}^3(4x_i - 2)\right)\right] - 1
\]
where $\Phi$ is the standard Gaussian CDF.  We use $d=4$ and set the failure 
threshold at $t=0$.  We define the input 
distribution as $x_i \stackrel{\mathrm{ind}}{\sim} \mathcal{N}(0.6, 0.11)$ for 
$i=1,\dots,4$, truncated to $[0, 1]$.

\section{Batch size comparison}\label{supp:batch}

Here we consider variations of the batch sizes used within the second
stage of our design.  When Stage 1 is complete and $n$-many evaluations
have been spent on contour location, the remaining budget ($B = N-n$) is to be
spent on the highest entropy samples from $X_M$.  Since the surrogate is used
to select these samples, there is an option to select them in smaller batches,
updating the surrogate with the new observations between each batch (indicated
by the dotted arrow in Figure \ref{fig:diagram}).  But smaller
batches require more computation as surrogate predictions must be obtained at
all $X_M$ locations after every batch update.

To investigate, we consider three different batch sizes for the Herbie function
example of Section \ref{sec:results}.  First, we choose all high entropy points in 
a single batch of size $B$ (this is the method used throughout the paper, thus
these results are duplicated from Figure \ref{fig:sims}).  Second, we consider
selecting one high entropy point at a time, updating the surrogate after every
high entropy point is observed to better inform the selection of subsequent high
entropy points.  Third, we consider selecting high entropy points in batches of size 10.
Note, changing the batch size used within Stage 2 has no
effect on Stage 1; each of these batch variations starts in the same place after
the completion of contour location and uses the same $X_M$.

\begin{figure}[!ht]
\centering
\includegraphics[width=8cm]{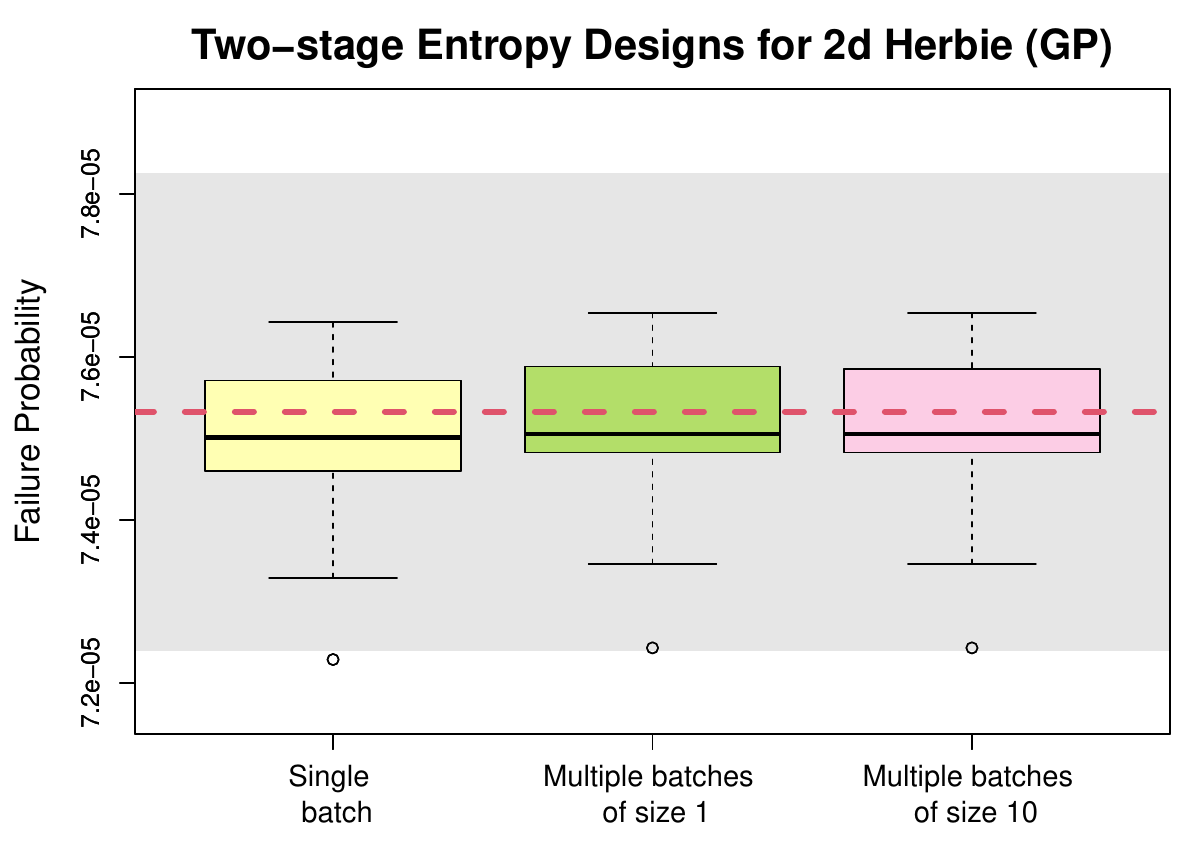}
\caption{Estimated failure probabilities shown over the true failure probability (dashed red)
$\pm 2$ standard errors (grey shading) for proposed two-stage designs of the Herbie function.
Variations use different batch sizes for selecting the highest entropy samples within Stage 2.}
\label{fig:batch}
\end{figure}

Figure \ref{fig:batch} shows the resulting failure probability estimates across 30
Monte Carlo repetitions.  There is no significant difference in performance
among the three batch sizes. The grey shaded region marks $\{\alpha\pm 2\sigma_\alpha\}$; 
it indicates the level of variability that we would expect from any Monte Carlo
estimator of specified $M$, even if every sample was classified accurately.  All 
methods consistently fall within this region, indicating effective estimation (note
the narrower y-axis range compared to Figure \ref{fig:sims}).  Given
the computational expense of repeatedly updating the surrogate, we embrace the single 
batch approach.

\end{document}